\documentclass[11pt,a4paper]{article}
\usepackage{cite}
\usepackage{graphicx}
\usepackage{amssymb}
\usepackage{amsmath}
\usepackage{amsfonts}
\usepackage{dsfont}
\usepackage{mathtools}
\usepackage{slashed}
\usepackage{rotating}
\usepackage{bbold,amsfonts}

\usepackage[utf8]{inputenc}
\usepackage{bm}
\usepackage{xcolor}
\usepackage{float}
\usepackage{braket}

\usepackage[height=8.8in,width=6.45in]{geometry}
\usepackage[font=small,labelfont=bf]{caption}
\usepackage[hidelinks]{hyperref}
\numberwithin{equation}{section}
\newcommand{\vev}[1]{{\left\langle #1 \right\rangle}}

\newcommand{\beq}{\begin{equation}}
\newcommand{\eeq}{\end{equation}}

\newcommand{\bsT}{\bar{\mathsf{T}}}
\newcommand{\sT}{\mathsf{T}}

\renewcommand{\a}{\alpha}
\renewcommand{\b}{\beta}

\renewcommand{\d}{\delta}
\newcommand{\pa}{\partial}
\newcommand{\g}{\gamma}
\newcommand{\G}{\Gamma}

\newcommand{\D}{\Delta}
\newcommand{\e}{\epsilon}

\renewcommand{\l}{\lambda}

\newcommand{\s}{\sigma}

\renewcommand{\t}{\tau}

\DeclareMathOperator{\Tr}{Tr}
\DeclareMathOperator{\tr}{tr}

\newcommand{\trp}{\Tr_{\cR}^\prime}

\newcommand{\ii}{\mathrm{i}}
\makeatletter
\newcommand*{\letterdef@}{}
\newcommand*{\letterdef}[3]{%
	\def\letterdef@##1{\expandafter\newcommand\csname #1\endcsname{#2{##1}}}%
	\@tfor\@tempa :=#3\do{\expandafter\letterdef@\expandafter{\@tempa}}}
\makeatother
\letterdef{c#1} {\mathcal}{ABCDEFGHIJKLMNOPQRSTUVWXYZ} 
\letterdef{rm#1}{\mathrm} {dDeimM} 
\newcommand{\tmb}[1]{{\mbox{\tiny{#1}}}}
\newcommand{\gym}{g_\tmb{YM}}
\newcommand{\cusp}{\text{cusp}}

\begin{document}
\begin{titlepage}
\vbox{
    \halign{#\hfil         \cr
           } 
      }  
\vspace*{15mm}
\begin{center}
{\LARGE \bf 
Emitted Radiation and Geometry
}

\vspace*{15mm}

{\Large L. Bianchi${}^{\,a}$, M. Bill\`o${}^{\,b,c}$, F. Galvagno${}^{\,b,c}$, A. Lerda${}^{\,d,c}$}
\vspace*{8mm}

${}^a$ Center for Research in String Theory - School of Physics and Astronomy\\
			Queen Mary University of London, Mile End Road, London E1 4NS, UK
			\vskip 0.3cm

${}^b$ Universit\`a di Torino, Dipartimento di Fisica,\\
			Via P. Giuria 1, I-10125 Torino, Italy
			\vskip 0.3cm
${}^c$ 
			I.N.F.N. - sezione di Torino,\\
			Via P. Giuria 1, I-10125 Torino, Italy 
			\vskip 0.3cm
${}^d$  Universit\`a del Piemonte Orientale,\\
			Dipartimento di Scienze e Innovazione Tecnologica\\
			Viale T. Michel 11, I-15121 Alessandria, Italy

\vskip 0.8cm
	{\small
		E-mail:
		\texttt{lorenzo.bianchi@qmul.ac.uk; billo,galvagno,lerda@to.infn.it}
	}
\vspace*{0.8cm}
\end{center}

\begin{abstract}
In conformal $\mathcal{N}=2$ Super Yang-Mills theory, the energy emitted by an accelerated heavy particle is computed by the one-point function of the stress tensor operator in the presence of a Wilson line. In this paper, we consider the theory on the ellipsoid and we prove a conjectured relation between the stress tensor one-point function and the first order expansion of the Wilson loop expectation value in the squashing parameter. To do this, we analyze the behavior of the Wilson loop for a small deformation of the background geometry and, at first order in the deformation, we fix the kinematics using defect CFT constraints. In the final part of the paper, we analyze the consequences of our results for the weak coupling perturbative expansion. In particular, comparing the weakly coupled matrix model with the ordinary Feynman diagram expansion, we find a natural transcendentality driven organization for the latter.

\end{abstract}
\vskip 1cm
	{
		Keywords: {$\mathcal{N}=2$ conformal SYM theories, Wilson loops, Brehmsstrahlung, rigid supersymmetry}
	}
\end{titlepage}

\tableofcontents
\vspace{1cm}

\section{Introduction}

The radiation emitted by an accelerated charged particle, often called Bremsstrahlung, is one of the most elementary physical observables in four-dimensional gauge theories. Despite this simplicity, examples of interacting quantum field theories where this quantity can be computed exactly are extremely rare. In classical electrodynamics the Larmor formula (and its relativistic generalization due to Li\'enard) predicts that the emitted energy is fully determined by the electric charge of the particle. An alternative way to derive the Larmor formula is to consider free Maxwell theory in four dimensions and compute the expectation value of a Wilson line. Free Maxwell theory in four dimensions has no scale, thus the straight Wilson line can be treated as a conformal defect. The emitted radiation is computed by slightly deforming the shape of the defect.

As usual, things get harder when one considers non-Abelian gauge theories, which are strongly coupled at low energies and where conformality is broken by quantum corrections. In light of this complexity, it is useful to restrict our attention to those examples of strongly interacting gauge theories which preserve conformality. These cases typically come with a larger symmetry group which includes supersymmetry, making them much more tractable.
For the maximally supersymmetric theory in four dimensions, $\mathcal{N}=4$ Super Yang Mills (SYM) theory, the combination of defect techniques and supersymmetric localization led to the derivation of a beautiful formula for the Bremsstrahlung function associated with the Maldacena Wilson loop \cite{Correa:2012at,Fiol:2012sg}, preserving half of the supercharges. Shortly after, the same result was confirmed by an integrability computation \cite{Correa:2012hh,Drukker:2012de}, providing one of the few examples of a quantity that is accessible to both techniques\footnote{It is worth mentioning that, before the achievement of these exact results, a huge effort has been made to compute the emitted radiation at strong coupling through the AdS/CFT correspondence \cite{Mikhailov:2003er,Athanasiou:2010pv,Hatta:2011gh,Fiol:2011zg}}. The work of \cite{Correa:2012at} heavily relied on the interpretation of the Wilson line as a superconformal defect. In particular, it was pointed out that the Bremsstrahlung function can be computed as the two-point function of an important defect operator, called the displacement operator. Furthermore, the same quantity can be related to the small angle limit of the cusp anomalous dimension, thus providing an interesting connection with massive scattering amplitudes.

Similar developments allowed to find an exact expression for the Bremsstrahlung function in ABJM theory \cite{Aharony:2008ug}, a three-dimensional relative of $\mathcal{N}=4$ SYM. In that case, two superconformal Wilson lines are known (see \cite{Drukker:2019bev} for a recent review). The immediate generalization of the Maldacena Wilson loop turns out to be $\frac16$BPS \cite{Drukker:2008zx,Chen:2008bp,Rey:2008bh} and its Bremsstrahlung function was already proposed in \cite{Lewkowycz:2013laa}. The maximally supersymmetric case \cite{Drukker:2009hy}, instead, involves also fermionic couplings and the computation of the exact Bremsstrahlung function required a long effort \cite{Forini:2012bb,Correa:2014aga,Bianchi:2014laa,Bianchi:2017ozk,Bianchi:2017svd} culminated in the closed-form expression presented in \cite{Bianchi:2018scb}. 

The crucial progress of \cite{Lewkowycz:2013laa} was the conjecture that the Bremsstrahlung function of $\mathcal{N}=4$ SYM and ABJM theory could be related to the one-point function of the stress tensor operator in the presence of the Wilson line. Motivated by this proposal and by strong perturbative evidence, the authors of \cite{Fiol:2015spa} extended this conjecture to the case of $\mathcal{N}=2$ conformal theories in four dimensions. In \cite{Bianchi:2018zpb} this relation was finally proven using supersymmetric Ward identities for the defect theory. In a further development, the authors of \cite{Fiol:2019woe} studied a different Lorentz invariant observable, called the invariant radiation rate, and they argued that, for conformal field theories, it can be universally related to the one-point function of the stress tensor -- see section \ref{Bandcusp} for a thorough discussion of this result.
It is then very clear, at present, that the crucial quantity for computing the emitted radiation in superconformal theories is the stress tensor one-point function.
The latter is fully determined by conformal invariance, up to an overall factor $h_W$, which depends non-trivially on the parameters of the theory:
\begin{equation}
\big\langle\,T_{00}(x)\,\big\rangle_W=\frac{h_W}{|x_{\perp}|^4}~.
\end{equation}
Here $|x_{\perp}|$ generically identifies the average orthogonal distance from the location of the line defect\footnote{For a fully consistent definition see section \ref{subsec:onepoint}.}.

In this paper we address the question of computing $h_W$ using supersymmetric localization. In particular, we present a derivation of the relation between $h_W$ and a small deformation of the geometric background, conjectured in \cite{Fiol:2015spa} (see also \cite{Mitev:2015oty,Gomez:2018usu}):
\begin{align}
	\label{conj1}
		\partial_b\ln \big\langle W_b\big\rangle\Big|_{b=1} =12\pi^2 h_W~.
\end{align}
Here $\big\langle W_b\big\rangle$ is the expectation value of the Wilson loop on the ellipsoid
with squashing parameter $b$ \cite{Hama:2012bg} and the value $b=1$ corresponds to the round sphere. The left hand side of this relation localizes and can be expressed in terms of a matrix model.

Our derivation only uses general properties of the geometric background and of defect CFTs, thus extending the relation \eqref{conj1} to any superconformal line defect. Furthermore, it provides a general recipe to extract exact results for the stress tensor one-point function by perturbing the background geometry. We stress that this is a peculiar feature of defect CFTs, where there is a non-vanishing one-point function and the first-order derivative gives a non-trivial result.

After proving the relation \eqref{conj1}, we carry out a careful analysis of the perturbative structure of the result, along the line of several results that have been achieved thanks to supersymmetric localization in the presence of a Wilson loop.
For the case of $\mathcal{N}=4$ SYM, the famous localization result for the $\frac12$BPS circular Wilson loops \cite{Erickson:2000af,Drukker:2000rr,Pestun:2007rz} can be extended to general configurations preserving less supersymmetry \cite{Zarembo:2002an,Drukker:2007dw,Drukker:2007qr} and it can be used to study correlators of the Wilson line with bulk local operators\cite{Giombi:2009ds,Bassetto:2009rt,Bassetto:2009ms,Giombi:2012ep,Bonini:2014vta}. More recently, an infinite family of defect CFT data was computed by considering a special class of defect operator insertions \cite{Giombi:2018qox,Giombi:2018hsx}. Moreover, thanks to maximal supersymmetry, various other techniques can be used to study these correlators, such as integrability \cite{Correa:2012hh,Drukker:2012de}, the conformal bootstrap \cite{Liendo:2018ukf} and the AdS/CFT correspondence \cite{Giombi:2017cqn,Beccaria:2019dws}.

Even though lowering the supersymmetry to $\mathcal{N}=2$ reduces the number of techniques at our disposal , supersymmetric localization can still compute the partition function of $\mathcal{N}=2$ Lagrangian theories on different geometries, and allows to capture several important observables; also the conformal bootstrap can still be used along the lines of \cite{Gimenez-Grau:2019hez}. For chiral primary operators, it turns out that the matrix model on $S^4$ contains all the necessary information to extract their two-point correlation functions \cite{Baggio:2014ioa,Gerchkovitz:2014gta,Baggio:2015vxa,Baggio:2016skg}, provided one solves the operator mixing induced by the map from the sphere to the plane \cite{Gerchkovitz:2016gxx,Rodriguez-Gomez:2016ijh,Rodriguez-Gomez:2016cem,Billo:2017glv,Bourget:2018obm, Bourget:2018fhe, Beccaria:2018xxl, Billo:2019job}. Moreover, in the presence of a circular Wilson loop the interacting matrix model captures the chiral primary one-point functions, at least in the conformal case \cite{Billo:2018oog,Beccaria:2018owt}, and its perturbative expansion suggests how to organize efficiently the loop corrections in the field theoretic evaluation of these observables \cite{Billo:2019fbi}.

The relation (\ref{conj1}), together with the series of equalities between the small angle limit of the cusp anomalous dimension, the displacement two-point function and the stress tensor one-point function discussed above, implies that all these apparently distinct observables are captured by the localization of a non-local operator on a deformed geometry. In particular, this provides a recipe to extract an exact prediction for a non-chiral scalar operator, such as the superprimary of the stress tensor multiplet.

Following the pattern that emerged in the case of chiral primary correlators, we study the constraints imposed by the matrix model expansion on the structure of the diagrams. We find that a limited class of diagrams contribute to the final result and that the matrix model provides a precious organizing principle, grouping different diagrams according to their color structure in a clever way.
In particular, the matrix model indicates that the lowest order contributions at fixed transcendentality arise from the loop corrections to a single propagator. We show that this structure is very natural, from the field theory side, 
when computing the cusp anomalous dimension and the displacement two-point function. It is instead far from being obvious in the case of the scalar superprimary expectation value, and could offer some general insight into the convenient way to approach perturbative computations involving non-chiral operators.

\section{Conformal SYM theories on the ellipsoid: a brief review}
\label{secn:ellipsoid}

We start by summarizing how to define  
$\mathcal{N}=2$ SYM theories on four-dimensional ellipsoids preserving rigid supersymmetry. We follow the analysis of \cite{Hama:2012bg}, whose conventions we largely adopt.

\subsection{The ellipsoid geometry}
\label{subsecn:ellips}

A four-dimensional ellipsoid can be defined as the surface in $\mathbb{R}^5$ described 
by the equation
\begin{equation}
	\label{defellipsoid}
		\frac{x_1^2+x_2^2}{\ell^2}+\frac{x_3^2+x_4^2}{\widetilde{\ell}^{\,2}}+\frac{x_5^2}{r^2}=1~.
\end{equation}
When $\ell=\widetilde\ell=r\equiv \mathsf{r}$, the ellipsoid becomes a round sphere $S^4$ of radius $\mathsf{r}$.
It is convenient to introduce the squashing parameter
\begin{equation}
b=\sqrt{\frac{\,\ell\,}{\,\widetilde{\ell}\,}}~,
\label{b}
\end{equation}
and use the following parametrization 
\begin{equation}
\label{parametrization}
\ell=l(b)\,b~,\quad \widetilde{\ell}=\frac{l(b)}{b}~,\quad
r=r(b)~,
\end{equation}
where $l(b)$ and $r(b)$ are such that $l(1)=r(1)=\mathsf{r}$. In this way, 
the limit $b\to 1$ corresponds to the sphere limit.

Following \cite{Hama:2012bg}, to describe the ellipsoid we adopt polar coordinates such that
\begin{equation}
\begin{aligned}
	\label{coords}
		x_1&=\ell\, \sin \rho \cos \theta \cos \varphi~,\\
		x_2&=\ell\, \sin \rho \cos \theta \sin \varphi~,\\
		x_3&=\widetilde{\ell}\,\sin \rho \sin \theta \cos \chi~,\\
		x_4&=\widetilde{\ell}\, \sin \rho \sin \theta \sin \chi~,\\
		x_5&=r\, \cos \rho~,
\end{aligned}
\end{equation}
where $\rho\in [0,\pi]$, $\theta\in [0,\pi/2]$, $\varphi\in [0,2\pi]$ and $\chi\in [0,2\pi]$. We collectively denote the polar coordinates as $\xi^\mu$, to distinguish them from the $\mathbb{R}^5$ coordinates $x_M$ (see Appendix \ref{app:Notations} for our conventions on indices). 

The ellipsoid metric $G_{\mu\nu}$ is simply given by the pullback of the flat Euclidean metric of the embedding space $\mathbb{R}^5$, namely
\begin{equation}
G_{\mu\nu}= \frac{\partial x_M}{\partial \xi^\mu}\,\frac{\partial x_N}{\partial \xi^\nu}\,\delta^{MN}
\label{metric}
\end{equation}
In our coordinate system, this metric is not diagonal and the corresponding vierbein $E^m=E^m_\mu d\xi^\mu$ are 
\begin{align}
	\label{vierbein}
		E^1 = \ell\, \sin \rho \cos \theta\, d\varphi~,\quad
		E^2 = \widetilde{\ell}\,\sin \rho \sin \theta\, d\chi~,\quad 
		E^3 = f \,\sin \rho\, d\theta+h\, d\rho~,\quad
		E^4 = g\, d\rho~,
\end{align}
with \cite{Hama:2012bg}
\begin{equation}
	\label{fhg}
		f =\sqrt{\ell^2\, \sin^2 \theta+\widetilde{\ell}^{\,2}\, \cos^2 \theta}~,~~~
		g = \sqrt{r^2\,\sin^2 \rho+ \frac{\ell^2\,\widetilde{\ell}^{\,2}}{f^2}\, \cos^2 \rho}~,~~~
		h = \frac{\widetilde{\ell}^{\,2}-\ell^2}{f}\,\cos \rho\, \sin \theta\, \cos \theta~.
\end{equation}
It is easy to see that $f\to \mathsf{r}$, $g\to \mathsf{r}$ and $h\to 0$ when $b\to 1$. Notice that since the polar coordinates $\xi^\mu$ are dimensionless, the metric $G_{\mu\nu}$ carries 
dimensions of $\mathrm{(length)}^2$; however, for the conformal invariant 
theories which we will consider, these dimensions can always be scaled away.

\subsection{Supersymmetric Lagrangians}
\label{subsecn:susy}

As shown in \cite{Hama:2012bg} following the general approach of \cite{Festuccia:2011ws}, 
in order to construct supersymmetric field theories on the ellipsoid it is necessary to introduce an off-shell (conformal) supergravity multiplet 
treated as a non-dynamical background (see also \cite{Klare:2013dka}).
In Euclidean signature, the fields of this supergravity multiplet, also called Weyl 
multiplet, are (see for example \cite{Freedman:2012zz})
\begin{equation}
	\label{offsm}
		G_{\mu\nu}~,\quad
		\psi^{\cI}_\mu~,\quad
		\sT_{\mu\nu}~,\quad 
		\bsT_{\mu\nu}~,\quad
		\widetilde{M}~,\quad
		\eta^{\cI}~,\quad
		V_{\mu}^0~,\quad
		(V_{\mu})^{\cI}_{\cJ}~,
\end{equation}
where $G_{\mu\nu}$ is the metric, $\psi^{\cI}_\mu$ (with $\cI=1,2$) is the gravitino, 
$\sT_{\mu\nu}$ and $\bsT_{\mu\nu}$ are, respectively, real self-dual and anti self-dual tensors\,%
\footnote{Do not confuse $\sT_{\mu\nu}$, written in an upright font, with the stress-energy tensor $T_{\mu\nu}$.}, $\widetilde{M}$ is a scalar field, $\eta^{\cI}$ is the dilatino, and finally 
$V_{\mu}^0$ and $(V_{\mu})^{\cI}_{\cJ}$ 
are the gauge fields of the  SO$(1,1)\times$SU$(2)_R$ R-symmetry. 

The action for a $\mathcal{N}=2$ SYM theory on an ellipsoid with squashing parameter $b$ has been
derived in \cite{Hama:2012bg} and is given by
\begin{equation}
\label{Sb}
S_b= \frac{1}{\gym^2}\,\int \!d^4\xi \,\sqrt{\det G}\,L
\end{equation}
where $L=L_{\tmb{YM}}+L_{\text{matter}}$. The first term,
${L}_{\tmb{YM}}$, accounts for the couplings of the gauge vector multiplet, which comprises the gauge connection $A_\mu$, the gaugino $\lambda_\cI$ and its conjugate
$\bar\lambda_\cI$, the scalar fields
$\phi$ and $\bar{\phi}$, and the auxiliary field $\cD_{\cI\cJ}$ -- all in the adjoint of the 
gauge group $\mathcal{G}$. 
The explicit expression of $L_{\tmb{YM}}$ is
\begin{equation}
\begin{aligned}
	\label{LYM}
 		L_{\tmb{YM}}&=\tr\bigg[\frac12 F^{\mu\nu} F_{\mu\nu}+ 16 F_{\mu\nu}(\bar \phi \mathsf{T}^{\mu\nu}+\phi \bar{\mathsf{T}}^{\mu\nu})+ 64\, \bar\phi^2 \mathsf{T}^{\mu\nu} \mathsf{T}_{\mu\nu}+ 64\, \phi^2 \bar{\mathsf{T}}^{\mu\nu} \bar{\mathsf{T}}_{\mu\nu}
 		-4 D_{\mu}\bar \phi D^{\mu} \phi
 		\\
 		&~~ +2 \Big(\widetilde{M}-\frac{R}{3}\Big) \bar \phi \phi -2 \ii \lambda^{\mathcal{I}}\s^{\mu} D_{\mu} \bar \lambda_{\mathcal{I}}- 2 \lambda^{\mathcal{I}} [\bar \phi,\lambda_{\mathcal{I}}]+ 2 \bar \lambda^{\mathcal{I}}[\phi,\bar \lambda_{\mathcal{I}}]+4 [\phi, \bar \phi]^2
 		-\frac{1}{2}\cD^{\cI\cJ}\cD_{\cI\cJ}\bigg]
\end{aligned}
\end{equation}
where $R$ is the Ricci scalar associated to the ellipsoid metric $G_{\mu\nu}$. 
Our conventions for the traces and the spinors are explained in Appendix~\ref{app:Notations}. Here we simply recall that the sum over repeated indices $\cI$ involves an $\epsilon$-tensor. For
example
\begin{equation}
\lambda^\cI \lambda_\cI = \epsilon^{\cI\cJ}\lambda_\cJ\lambda_\cI
\end{equation}
with $\epsilon^{12}=1$.

A few comments are in order. Following \cite{Hosomichi:2016flq}, we have written the coefficient of the $\bar\phi \phi$-term as twice $\big(\widetilde{M}-\frac{R}{3}\big)$. This combination is equivalent to the field $M$ used in \cite{Hama:2012bg}, but for our purposes it is more convenient to distinguish the contribution due the background field $\widetilde{M}$ from the one due to the curvature. 
Moreover, if we add the
$R\,\bar{\phi}\phi$-term to the scalar kinetic term, we obtain 
\begin{equation}
-4\tr\Big(D_{\mu}\bar \phi D^{\mu} \phi+ \frac{R}{6}\,\bar\phi \phi\Big)~.
\end{equation}
The coefficient of $1/6$ in front of the curvature shows that the scalar fields of the vector multiplet are conformally coupled to the ellipsoid metric. We also note that the $\mathrm{SU}(2)_R$ connection 
$(V_{\mu})^{\cI}_{\cJ}$ does not appear explicitly in the Lagrangian, but only 
through the covariant derivative of the gaugino, which is defined as 
\begin{align}
	\label{gaugcovder}
		D_{\mu} \bar \lambda^{\dot \a}_{\mathcal{I}} 
		=\pa_{\mu} \bar \lambda^{\dot \a}_{\mathcal{I}}-\ii [A_{\mu}, \bar \lambda^{\dot \a}_{\mathcal{I}}]+ \frac14 \omega_{\mu}^{mn} (\bar \s_{mn})^{\dot \a}{}_{\dot \b} \bar \lambda^{\dot \b}_{\mathcal{I}}+ \ii \bar \lambda^{\dot \a}_{ \mathcal{J}} (V_{\mu})^{\mathcal{J}}{}_{\mathcal{I}}~,		
\end{align}
where $\omega_{\mu}^{mn}$ is the spin-connection, and similarly for the left-handed components.
Note that the gauge field $V_{\mu}^0$ has been set to zero, as in \cite{Hama:2012bg}. We discuss this choice at the end of this subsection.

The matter part of the Lagrangian, $L_{\text{matter}}$, accounts for the couplings of $\cN=2$ 
hypermultiplets transforming in a (generically reducible) representation $\mathcal{R}$ of the gauge group. The number of these hypermultiplets is clearly equal to the dimension of $\mathcal{R}$, which we denote simply by $r$. If the index $i_{\mathcal{R}}$ of the matter representation equals that of the adjoint, then the resulting $\cN=2$ SYM theory is conformal\,%
\footnote{For example, for SU($N$) if $\mathcal{R}$ is the sum of $N_f$ fundamental representations, each of dimension $N$ and index $1/2$, we have $r=N_f\,N$, $i_\mathcal{R}=N_f/2$ and the condition for conformal invariance is the familiar constraint $N_f/2=N$.}. In the following we will restrict to this case.
If we denote the scalar fields of the hypermultiplets by $q_{\cI\cA}$ and their fermionic
partners by $\psi_{\cA}$ and $\bar\psi_\cA$, with $\cA=1,\ldots,2r$ being an index of Sp($r$),
the matter Lagrangian takes the form
\begin{equation}
\begin{aligned}
	\label{Lhm} 	 
	 	L_{\text{matter}}&=\frac12 D_{\mu} q^\cI D^{\mu} q_\cI- q^\cI\{\phi, \bar \phi\} q_\cI
	 	-\frac18 q^\cI q_\cI q_\cJ q^\cJ+\frac18 \Big(\widetilde{M}-\frac{2}{3}R\Big)q^\cI q_\cI\\
 		&~~-\frac{\ii}{2} \bar \psi \bar \sigma^{\mu} D_{\mu} \psi -\frac12 \psi \phi \psi +\frac12 \bar \psi \bar \phi \bar \psi+\frac{\ii}{2}\psi \s^{\mu\nu} \mathsf{T}_{\mu\nu} \psi -\frac{\ii}{2}
 		\bar \psi \bar \s^{\mu\nu} \bar{\mathsf{T}}_{\mu\nu} \bar \psi- q^\cI \lambda_\cI \psi+ \bar \psi \bar \lambda_\cI q^\cI ~.
\end{aligned}
\end{equation}
Here the sum over the Sp($r$) indices has been understood. If one wants to write it explicitly, one
has for example
\begin{align}
 q^{\mathcal{I}}  q_{\mathcal{I}}= \Omega^{\mathcal{A}\mathcal{B}}\,
 q^{\mathcal{I}}_{~\mathcal{B}}\,q_{\mathcal{I}\mathcal{A}} ~,
\end{align}
where $\Omega^{\mathcal{A}\mathcal{B}}$ is the real anti-symmetric invariant tensor of Sp($r$).
Notice that the matter fields are coupled to the vector multiplet through an embedding of the gauge group into $\mathrm{Sp}(r)$ and that, as before, 
the $\mathrm{SU}(2)_R$ connection appears only in the covariant derivatives defined by
\begin{align}
	\label{covderpsi}
		D_{\mu} q_{\mathcal{I} \mathcal{A}}&=\pa_{\mu} q_{\mathcal{I} \mathcal{A}}-\ii (A_{\mu})_{\mathcal{A}}{}^{\mathcal{B}} q_{\mathcal{I} \mathcal{B}}+ \ii q_{\mathcal{J}\mathcal{A}} (V_{\mu})^{\mathcal{J}}{}_{\mathcal{I}}~.
\end{align}
Again, in the Lagrangian (\ref{Lhm}) we have replaced the scalar $M$ appearing in \cite{Hama:2012bg} with $\big(\widetilde{M}-\frac{R}{3}\big)$ in order to disentangle the contribution due to
the curvature from that due to the scalar field of the supergravity multiplet. Moreover, combining the
$R\,q^\cI q_\cI$-term with the kinetic terms we obtain
\begin{equation}
\frac12 \Big(D_{\mu} q^\cI D^{\mu} q_\cI+\frac{R}{6}\,q^\cI q_\cI\Big)
\end{equation}
which shows that also the scalar fields of the matter hypermultiplets are conformally coupled to
the curvature of the ellipsoid.

The action $S_b$ in (\ref{Sb}) is invariant under the $\cN=2$ supersymmetry transformations 
of the gauge and matter fields given in Appendix \ref{app:SUSYtransf} provided the supergravity
background is carefully chosen. In particular, the metric $G_{\mu\nu}$ must be that of the ellipsoid
as in (\ref{metric}), while $\sT_{\mu\nu}$, $\bar{\sT}_{\mu\nu}$, $\widetilde{M}$ 
and $(V_{\mu})^{\cI}_{\cJ}$ must assume background values determined by solving 
the Killing spinor equations that ensure the vanishing of the supersymmetry transformations of the gravitino and dilatino. Their expressions, found in \cite{Hama:2012bg} and to be recalled below, depend on the geometric properties of the ellipsoid, and in particular on the squashing 
parameter $b$. 
As already mentioned, the SO$(1,1)_R$ connection $V_{\mu}^0$ can be consistently set to zero, since the Killing spinor equations determine the background geometry up to some residual degrees of freedom. This choice pursued in \cite{Hama:2012bg} is justified also by the necessity of reproducing the so-called $\Omega$-background \cite{Nekrasov:2002qd} at the North and South poles of the ellipsoid and is allowed by a residual symmetry from the 
supersymmetry conditions, as widely explained in \cite{Klare:2013dka}. 

We conclude by observing that on a round sphere with no background
fields turned on, except for the metric, the above actions reduce to those considered for the
localization on $S^4$ in \cite{Pestun:2007rz}.

\subsection{Supergravity background}
\label{subsecn:background}
The Killing spinor equations provide specific geometric constraints that allow to fix the 
profile of the background fields, although not uniquely. In \cite{Hama:2012bg}  it was found
that these fields are given by\footnote{To be precise \cite{Hama:2012bg} contains the explicit expression of $M$, not $\widetilde{M}$. To obtain the latter, one can simply use the relation $\widetilde{M}=M+\frac{R}{3}$ and the Ricci curvature associated to the metric \eqref{metric}, $R=3\left(\frac{1}{g^2}+\frac{r^2}{f^2g^2}\right)$.}
\begin{subequations}
\label{HHsol}
\begin{align}	
	\widetilde M & =\frac{1}{f^2}+\frac{h^2+r^2}{f^2 g^2}-\frac{4}{fg}
		+ \Delta \widetilde M~,\label{Msol}\\[2mm]
		\mathsf{T}_{\a}{}^{\b} & = \frac14\Big( \frac{1}{f}-\frac{1}{g}\Big) (\t_{\theta}^1)_{\a}{}^{\b} 
		+\frac{h}{4 f g} (\t_{\theta}^2)_{\a}{}^{\b} + \Delta \mathsf{T}_{\a}{}^{\b}~,\label{Tsol}\\[2mm]
		\bar{\mathsf{T}}^{\dot \a}{}_{\dot \b} & = \frac14\Big( \frac{1}{f}-\frac{1}{g}\Big) 
		(\t_{\theta}^1)^{\dot \a}{}_{\dot \b} -\frac{h}{4 f g} (\t_{\theta}^2)^{\dot \a}{}_{\dot \b} 
		+ \Delta \bar{\mathsf{T}}^{\dot \a}{}_{\dot \b}~,\label{barTsol}
		\end{align}
\end{subequations}
where the functions $f$, $g$ and $h$ are defined in (\ref{fhg}), while the matrices $\t^i_\theta$ are
\begin{align}
	\label{tthetamat}
		\t^i_\theta = \t^i \,\begin{pmatrix} \rme^{+\ii \theta} &0  \\ 0& \rme^{-\ii \theta} 
		\end{pmatrix} ~,
\end{align}
with $\t^i$ being the usual Pauli matrices. Note that the self-dual and anti self-dual
tensors $\mathsf{T}_{\mu\nu}$ and $\bar{\mathsf{T}}_{\mu\nu}$ are related to the matrices 
$\mathsf{T}_{\a}{}^{\b} $ and $\bar{\mathsf{T}}^{\dot \a}{}_{\dot \b} $ in 
(\ref{Tsol}) and (\ref{barTsol}) according to
\begin{align}
	\label{Tab}
		\mathsf{T}_{\a}{}^{\b} = -\ii\left(\sigma^{\mu\nu}\right)_{\a}{}^{\b}\, \mathsf{T}_{\mu\nu}~,~~~
		\bar{\mathsf{T}}^{\dot \a}{}_{\dot \b} =-\ii \left(\bar{\sigma}^{\mu\nu}\right)^{\dot \a}{}_{\dot \b} \,\bar{\mathsf{T}}_{\mu\nu}~.
\end{align}
Finally, in each line of (\ref{HHsol}) the last contribution, indicated with a $\Delta$, depends 
on three arbitrary functions $c_1$, $c_2$ and $c_3$, which parameterize the ambiguity of the background solution. In fact we have \cite{Hama:2012bg}
\begin{equation}
\begin{aligned}
	\label{DeltaM}
		\D \widetilde{M}&= 8\,\Big(\frac{1}{g} \pa_{\rho}- \frac{h}{g f \sin \rho} \pa_{\theta} + \frac{\ell^2 \tilde \ell^2 \cos \rho}{g f^4 \sin \rho}+ \frac{(\ell^2+\tilde{\ell}^2-f^2)\cos \rho}{g f^2 \sin \rho}- \frac{\cos \rho}{f \sin \rho} \Big) c_1\\[1mm]
		&~~+8\,\Big(\frac{1}{f\sin \rho}\pa_\theta+ \frac{\ell^2 \tilde{\ell}^2 h\,\cos \rho}{g^2 f^4 \sin \rho} + \frac{2 \cot 2\theta}{f \sin \rho}- \frac{h \cos \rho}{f g \sin \rho} \Big) c_2 -16 (c_1^2+c_2^2+c_3^2)~,
\end{aligned}
\end{equation}
and
\begin{subequations}
\label{DeltaTs}
\begin{align}
		\label{DeltaT}
		\D  \mathsf{T}_{\a}{}^{\b} &= \tan \frac{\rho}{2} \,\Big(c_1  (\t_{\theta}^1)_{\a}{}^{\b}+c_2  (\t_{\theta}^2)_{\a}{}^{\b}+c_3  (\t^3)_{\a}{}^{\b} \Big)~,\\[2mm]
		\label{DeltabarT}
		\D  \bar{\mathsf{T}}^{\dot \a}{}_{\dot \b} &= \cot \frac{\rho}{2} \,\Big(\!-c_1  (\t_{\theta}^1)^{\dot \a}{}_{\dot \b}+c_2  (\t_{\theta}^2)^{\dot \a}{}_{\dot \b}+c_3  (\t^3)^{\dot \a}{}_{\dot \b}\Big)~.
\end{align}
\end{subequations}
It is easy to check that in the sphere limit when $b\to 1$, all non $\Delta$-terms in (\ref{HHsol})
vanish. Therefore, since on the sphere the only surviving background field
is the metric, we must require that also $\Delta\widetilde{M}$, $\Delta \mathsf{T}$ and
$\Delta \bar{\mathsf{T}}$ vanish when $b=1$. In turn this requirement implies the $c_i$'s must be
zero at $b=1$, {\it{i.e.}} they must have the following form
\begin{align}
c_i= c'_i (b-1) + {O}\big((b-1)^2\big)~.
\label{ci}
\end{align}
It is important to remark that also the SU$(2)_R$ connection $(V_{\mu})^{\cI}_{\cJ}$ acquires a
background profile in the supersymmetric realization of the SYM theory on the ellipsoid; however,
in Section \ref{subsec:onepoint} we will explain why the explicit expression of this profile is not needed in the present work and for this reason we have not reported it here.

\section{Relating $h_W$ to the ellipsoid deformation} 
\label{secn:WL}
In the set-up reviewed in the previous section, we want to analyze how the vacuum expectation values of gauge invariant operators in the conformal $\cN=2$ SYM theory respond to a deformation of the ellipsoid geometry, and specifically how they depend on the squashing parameter $b$ in the
vicinity of the sphere limit. The goal is to find a direct relation between the quantity $h_W$ defined in the introduction and the vacuum expectation value of half-BPS Wilson loops to prove the conjecture (\ref{conj1}).

Let us consider a gauge invariant operator $X_b$ which may depend on the ellipsoid squashing parameter. Its vacuum expectation
value is
\begin{equation}
\big\langle X_b\big\rangle =\frac{1}{Z_b}\,\int \!D A \,\, \rme^{-S_b}\, X_b~,
\label{vevO}
\end{equation}
where $A$ here denotes schematically all fields in the conformal $\cN=2$ SYM theory whose
action $S_b$ is given in (\ref{Sb}), and $Z_b$ is the partition function
\begin{equation}
Z_b=\int \!D A \,\, \rme^{-S_b}~.
\label{Zb}
\end{equation}
{From} this definition it easily follows that
\begin{equation}
\begin{aligned}
\partial_b\,\ln \big\langle X_b\big\rangle\Big|_{b=1} &=
\frac{-\big\langle \partial_bS_b\,X_b\big\rangle
+\big\langle \partial_bS_b\big\rangle\,\big\langle X_b\big\rangle +\big\langle
\partial_b X_b\big\rangle}{\big\langle X_b\big\rangle}\Big|_{b=1}\\[2mm]
&=
-\frac{\big\langle \!:\!\partial_b S_b\!: X_b\big\rangle}{\big\langle X_b\big\rangle}\,
\Big|_{b=1}
+
\frac{\big\langle
\partial_b X_b\big\rangle}{\big\langle X_b\big\rangle}\,\Big|_{b=1}
\end{aligned}
\label{dbO}
\end{equation}
where the $:\,:$'s indicate the normal ordering, namely the subtraction of all possible
self-interactions. This expression should not depend on the parametrization (\ref{parametrization})
of the scales of the ellipsoid.

Since the action $S_b$ depends on $b$ only through the background supergravity fields, we have
\begin{equation}
\label{dbSbop}
\begin{aligned}
		\partial_b S_b
		& = \int \!d^4 \xi\,\sqrt{\det G}  \bigg[\frac{1}{\sqrt{\det G}}\frac{\pa (\sqrt{\det G}\,L) }{\pa G^{\mu\nu}}\, \pa_b G^{\mu\nu}+
		\frac{\pa L }{\pa (V^{\mu})^{\mathcal{J}}{}_{\mathcal{I}}}
		\,\pa_b (V^{\mu})^{\cJ}{}_{\cI}\\[2mm]
		&\qquad~~+ \frac{\pa L }{\pa \mathsf{T}^{\mu \nu}} \,\pa_b \mathsf{T}^{\mu \nu}
		+  \frac{\pa L }{\pa \bar{\mathsf{T}}^{\mu \nu}} \,
		\pa_b \bar{\mathsf{T}}^{\mu \nu}
		+  \frac{\pa L}{\pa \widetilde{M}} \,\pa_b 
		\widetilde{M}\bigg]~. 
\end{aligned}
\end{equation}
We are interested in evaluating this expression at $b=1$.
By definition, the variation of the action  with respect to the metric at $b=1$
yields the stress-energy tensor $T_{\mu\nu}$ on the sphere. More precisely, we have:
\begin{align}
	\label{defT}
		\frac{\pa (\sqrt{\det G}\,L) }{\pa G^{\mu\nu}}\,\Big|_{b=1} 
		= - \frac 12 \sqrt{\det G^0}\,\,T_{\mu\nu}
\end{align}
where $G^0_{\mu\nu}$ is the metric on the round sphere $S^4$, namely
\begin{equation}
G^0_{\mu\nu}= \lim_{b\to1} G_{\mu\nu}~.
\label{G0}
\end{equation}
Similarly, the variations of the action with respect to the other background 
fields of the supergravity multiplet 
yield the other bosonic components of the stress-energy tensor supermultiplet, known also as the supercurrent multiplet. With the conventions given in Appendix~\ref{app:SUSYtransf}, we have
\begin{equation}
\label{defoperators}
\begin{aligned}
\frac{\pa L }{\pa (V^{\mu})^{\mathcal{J}}{}_{\mathcal{I}}}\,\Big|_{b=1} 
		&= -\frac{\ii}{2} (t_{\mu})_{\mathcal{J}}{}^{\mathcal{I}}~,~~~~~~~~ &
\frac{\pa L }{\pa \mathsf{T}^{\mu \nu}}\,\Big|_{b=1} 
		&= -16H_{\mu\nu}~,\\[2mm]
\frac{\pa L }{\pa \bar{\mathsf{T}}^{\mu \nu}}\,\Big|_{b=1} 
		&= -16\bar H_{\mu\nu}~, &
\frac{\pa L }{\pa \widetilde{M}}\,\Big|_{b=1} 
	     &= -O_2~.
\end{aligned}
\end{equation}
Using the Lagrangian $L = L_{\tmb{YM}} +  L_{\text{matter}}$ reviewed in the previous section, we find 
\begin{equation}
\label{varscmultiplet}
\begin{aligned}
(t_{\mu})_{\mathcal{J}}{}^{\mathcal{I}}&= 
		 4 \ii \tr[\l^{\mathcal{I}} \s^{\mu} \bar \lambda_{\mathcal{J}}]-2\ii \tr[\l^{\mathcal{K}} \s^{\mu} \bar \lambda_{\mathcal{K}}]\,\delta_{\mathcal{J}}^{\mathcal{I}}
		+ q^{\mathcal{I}}\overset{\leftrightarrow}{D}_{\mu} q_{\mathcal{J}}+\frac{1}{2} q^{\mathcal{K}}\overset{\leftrightarrow}{D}_{\mu} q_{\mathcal{K}}\,\delta_{\mathcal{J}}^{\mathcal{I}}~,\\
		H_{\mu\nu}&= -\tr[F^+_{\mu\nu} \,\bar \phi]-\frac{\ii}{32} \psi \s_{\mu\nu} \psi~,\\
		{\bar H}_{\mu\nu}&=-\tr[F^-_{\mu\nu} \,\phi]+\frac{\ii}{32} \bar \psi \bar \s_{\mu\nu} \bar \psi ~,\\
		O_2&= -2 \tr[\bar \phi \phi] -\frac18 q^{\mathcal{I}} q_{\mathcal{I}}
\end{aligned}
\end{equation}
where $F_{\mu\nu}^+$ and $F_{\mu\nu}^-$ are the self-dual and anti self-dual parts of the
gauge field strength.
As a matter of fact, in the following we will not really need these explicit expressions, but we quoted them here to allow the check that the coefficients relating them to the variations of the Lagrangian as given in (\ref{defoperators}) are consistent with the supersymmetry transformations 
reported in Appendix~\ref{app:SUSYtransf} -- indeed, these coefficients will be important for our results. 

With these definitions, we can rewrite (\ref{dbSbop}) as
\begin{equation}
\begin{aligned}
		\partial_b S_b\,\big|_{b=1}&=- \int \!d^4 \xi \,\sqrt{\det G^0} \,
		\bigg[\frac{1}{2}\,T_{\mu\nu}\, \pa_b G^{\mu\nu}\big|_{b=1}+\frac{\ii}{2}\,          
		(t_{\mu})_{\mathcal{J}}{}^{\mathcal{I}}
		\,\pa_b (V^{\mu})^{\cJ}{}_{\cI}\big|_{b=1}
		\\[2mm]
		&\qquad~~+ 16 H_{\mu\nu}\,\pa_b \mathsf{T}^{\mu \nu}\big|_{b=1}
		+ 16 \bar{H}_{\mu\nu}\,
		\pa_b \bar{\mathsf{T}}^{\mu \nu}\big|_{b=1}
		+ O_2 \,\pa_b 
		\widetilde{M}\big|_{b=1}\bigg]~. 
\end{aligned}
\label{partialSb}
\end{equation}
In the following we will use this set-up to study how a half-BPS Wilson loop responds to
a deformation of the ellipsoid.

\subsection{Half-BPS Wilson loops}
\label{subsecn:WL}

On the ellipsoid there are two possible half-BPS Wilson loop defects. 
One wraps the circle of radius $\ell$ in the $x^1,x^2$ plane, the other wraps the circle of radius 
$\tilde\ell$ in the $x^3,x^4$ plane. The two configurations can be exchanged by sending
$b \leftrightarrow 1/b$. 
\begin{figure}[H]
	\begin{center}
		\includegraphics[scale=0.7]{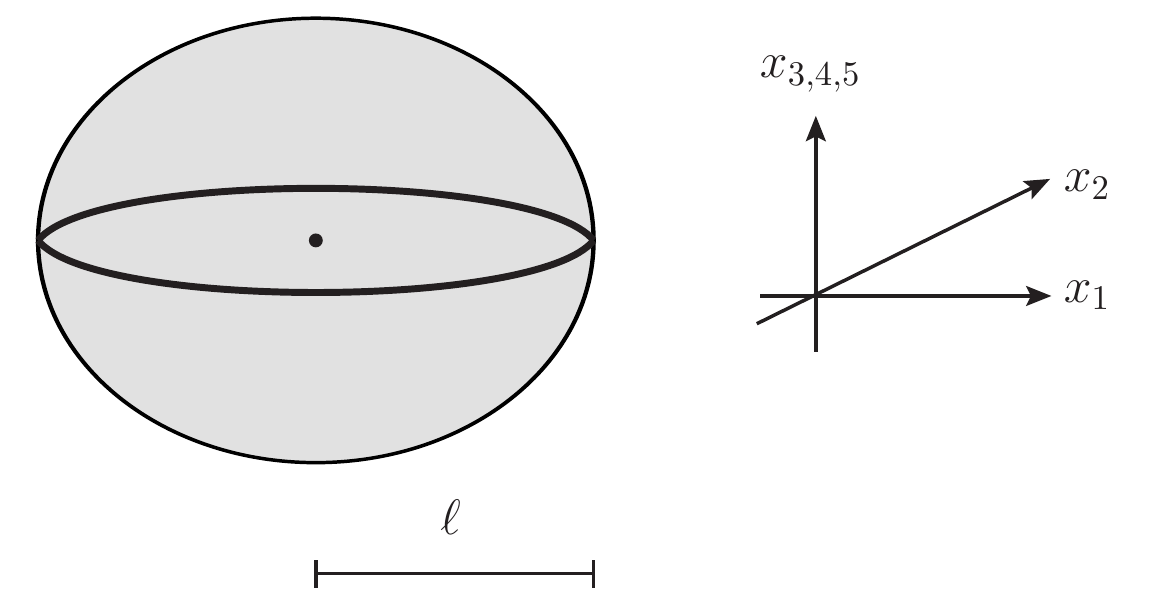}
	\end{center}
	\caption{Wilson loop wrapped around the circle of radius $\ell$ in the $x^1,x^2$ plane on the ellipsoid.}
	\label{fig:ellipsoid}
\end{figure}
Without loss of generality we can choose to wrap the circle of radius $\ell$, see 
Fig.~\ref{fig:ellipsoid}. Hence, in the polar coordinates (\ref{coords}), the
Wilson loop locus $\mathcal{C}$ is defined by $\chi=\theta=0$, $\rho=\pi/2$. The explicit expression of
this Wilson loop is \cite{Hama:2012bg}
\begin{equation}
\label{WL}
W_b =\frac{1}{d_\mathcal{R}}\,\tr_{\mathcal{R}}\,\mathcal{P}
\exp\bigg[\ii\!\int_\mathcal{C}	\!d\varphi\,\Big(A_\varphi-\ell(\phi+\bar\phi)\Big)\bigg]
\end{equation}
where $d_\mathcal{R}$ is the dimension of the representation $\mathcal{R}$ in which the
Wilson loop transforms. Notice that this operator may explicitly depend on $b$ through the
coefficient $\ell$ of the scalar part, once the parametrization (\ref{parametrization}) is used.

{From} the formul\ae~(\ref{dbO}) and (\ref{partialSb}), we obtain
\begin{equation}
\begin{aligned}
\partial_b\ln \big\langle W_b\big\rangle\Big|_{b=1} &=
\int \!d^4 \xi \,\sqrt{\det G^0} \,
		\bigg[\frac{1}{2}\,\big\langle T_{\mu\nu}\big\rangle_W \, \pa_b G^{\mu\nu}\big|_{b=1}+\frac{\ii}{2}\, \big\langle
		(t_{\mu})_{\mathcal{J}}{}^{\mathcal{I}}\big\rangle_W
		\,\pa_b (V^{\mu})^{\cJ}{}_{\cI}\big|_{b=1}
		\\[2mm]
		&\qquad~~+ 16 \,\big\langle H_{\mu\nu}\big\rangle_W
		\,\pa_b \mathsf{T}^{\mu \nu}\big|_{b=1}
		+ 16\, \big\langle \bar{H}_{\mu\nu}\big\rangle_W \,
		\pa_b \bar{\mathsf{T}}^{\mu \nu}\big|_{b=1}\\[2mm]
		&\qquad~~+ \big\langle O_2\big\rangle_W \,\pa_b 
		\widetilde{M}\big|_{b=1}\bigg]
		+
\frac{\big\langle
\partial_b W_b\big\rangle}{\big\langle W_b\big\rangle}\,\Big|_{b=1}
\end{aligned}
\label{dlogW}
\end{equation}
where we have adopted the short-hand notation $\big\langle X \big\rangle_W$ to denote the normalized one-point function of $:\!X\!:$ in the presence of the Wilson loop on the sphere, namely
\begin{equation}
\big\langle X \big\rangle_W \equiv \frac{\big\langle \!:\!X\!: W_b \big\rangle}{\big\langle W_b \big\rangle}
\Big|_{b=1} =\frac{\big\langle X W \big\rangle}{\big\langle W \big\rangle}
-\big\langle X  \big\rangle 
\label{XW1}
\end{equation}
with $W$ denoting the Wilson loop on the sphere.
Our goal is to explicitly calculate the integrals in (\ref{dlogW}). 

\subsection{Non-vanishing one-point functions}
\label{subsec:onepoint}

The half-BPS Wilson line in a $\mathcal{N}=2$ SCFT preserves an 
$\mathfrak{osp}(4^*|2)$ sub-algebra of the full $\mathfrak{su}(2,2|2)$ superconformal algebra and, in particular, it preserves the one-dimensional conformal group. 
In a defect conformal field theory, the functional form of the one-point functions of bulk operators  is entirely fixed by the preserved defect 
(super)-conformal symmetry. To write their expressions for the conformal SYM theory we
are considering, it is convenient to resort to the so-called embedding formalism (see for example \cite{Billo:2016cpy}).

\paragraph{Embedding coordinates:} We introduce light-cone embedding coordinates $P^\cM=(P^0,P^M)$, with $\cM=0,\dots,5$ and $M=1,\dots,5$. The metric in the six-dimensional space is the Minkowski metric, $\eta_{\cM\cN}=\text{diag}(-1,1,1,1,1,1)$. 
Different sections of the light-cone lead to different expressions in real space, all of which are related to each other by a conformal transformation. Since the sphere is conformally equivalent to a plane, we can choose a specific light-cone section. In particular, for a 
one-dimensional defect like our Wilson loop, 
we distinguish the coordinates $P^\cM$ into three parallel ones (associated to the residual SO$(1,2)$ conformal group of the defect) and three orthogonal ones (associated to the SO$(3)$ orthogonal rotations). Of course, when the defect sits on a sphere, like in our case, the conformal Killing vectors are not as immediate as in the planar case, but nevertheless there is a very natural choice to make
for the light-cone section, namely
\begin{align}
\label{P}
P^\cM=\big(\mathsf{r},x_M\big|_{b=1}\big)
\end{align}
with $x_M\big|_{b=1}$ are the coordinates given in (\ref{coords}) evaluated 
on the sphere of radius $\mathsf{r}$.
The coordinate $P^0$ is determined by the condition 
$P^\cM\eta_{\cM\cN}P^\cN=0$, while the two coordinates along which the defect stretches, 
{\it{i.e.}} $x_1$ and $x_2$, are the parallel coordinates in embedding space. 
To sum up, in our case $P^0=\mathsf{r}$ and $P^{1,2}=x_{1,2}\big|_{b=1}$ are the 
parallel coordinates, while $P^{3,4,5}=x_{3,4,5}\big|_{b=1}$ are the orthogonal ones. 
With this assignment, the extraction of the orthogonal and parallel scalar products, denoted
respectively by $\circ$ and $\bullet$, is a trivial exercise:
\begin{equation}
\begin{aligned}
P \circ P &=\big( x_3^2+x_4^2+x_5^2\big)\big|_{b=1}=\mathsf{r}^2\big(\cos^2\rho+
\sin^2\theta\sin^2\rho\big)~,\\[1mm]
 P \bullet P &= -\mathsf{r}^2+\big(x_1^2+x_2^2\big)\big|_{b=1}=-\mathsf{r}^2\big(1-\cos^2\theta\sin^2\rho \big)=-P\circ P~.
\end{aligned}
\label{PP}
\end{equation}

A further ingredient that is needed to write the expression of the one-point functions
is the projection of indices using the so-called $z$-variables \cite{Billo:2016cpy, Lauria:2018klo}.
For a symmetric traceless tensor, like the stress-energy tensor $T^{\mu\nu}$, one can contract 
all indices with a complex vector $z_\mu$, 
such that $z\cdot z\,\equiv \, z^\mu G^0_{\mu\nu} z^{\nu}=0$. 
Then the one-point function of this tensor in the presence of a defect is a polynomial in $z$. 
If one needs the one-point function with open indices, one can apply to this polynomial
the Todorov operator \cite{Dobrev:1975ru}
\begin{align}
\label{todorov}
\cD_{\mu}=\Big(1+z\cdot \frac{\pa}{\partial z} \Big)\frac{\partial}{\partial z^{\mu}}-\frac12 z_{\mu} \frac{\partial^2}{\partial z \cdot \partial z}~.
\end{align}
A useful, but not unique, strategy to extend this prescription to the light-cone is to introduce
a vector $Z$ in the embedding space given by
\begin{align}\label{Zlightcone}
Z^{\cM}=z^{\mu} \pa_\mu P^\cM~.
\end{align}
Using the relation
\begin{align}
\pa_\mu P^\cM\,\eta_{\cM\cN} \,\pa_\nu P^\cN=G^0_{\mu\nu}~,
\end{align}
which can be easily verified in our case, one can check that
\begin{equation}
P^\cM\eta_{\cM\cN} Z^\cN=Z^\cM\eta_{\cM\cN} Z^\cN=0
\end{equation}
if $z\cdot z=0$. 

For tensors that are not symmetric or traceless, the procedure is a bit more intricate and a
general discussion can be found in \cite{Lauria:2018klo}. For our purposes, it is enough however 
to consider the case of the anti-symmetric two-index tensors. In this case two different $z$-vectors,
$z^{(1)}$ and $z^{(2)}$, are introduced and the one-point function is expressed as a polynomial in these two vectors. Then, by introducing two $Z$-vectors in the embedding space following the same steps as in (\ref{Zlightcone}), one can easily extend this formalism to the light-cone and obtain the explicit form of the one-point function.

\paragraph{The relevant one-point functions:}
In the presence of a conformal line defect, only operators with even spin can acquire an expectation value \cite{Billo:2016cpy} (the situation may be different for special cases where parity odd structures are available, but this is not the case for a line defect in four dimensions). Therefore, in
our case, the one-point function of $(t_{\mu})_{\mathcal{J}}{}^{\mathcal{I}}$ vanishes:
\begin{equation}
\big\langle (t_{\mu})_{\mathcal{J}}{}^{\mathcal{I}}\big\rangle_W=0~,
\label{vevt}
\end{equation}
and the only non-zero one-point functions are those of the stress-tensor $T_{\mu\nu}$, of the two anti-symmetric tensors $H_{\mu\nu}$ and $\bar{H}_{\mu\nu}$, and the scalar operator $O_2$.

The one-point function of the stress-energy tensor can be extracted from \cite{Billo:2016cpy}
and reads
\begin{equation}
\label{zzT0}
z^{\mu} z^{\nu} \big\langle T_{\mu\nu}\big\rangle_W
= 4 h_W \frac{(P\circ Z)^2 - (Z\circ Z)\,(P\circ P)}{(P\circ P)^3}
\end{equation}
where $h_W$ is the same quantity discussed in the introduction. Using the explicit expressions of $P$ and $Z$ given in (\ref{P}) and (\ref{Zlightcone}), we find
\begin{align}
z^{\mu} z^{\nu} \big\langle T_{\mu\nu}\big\rangle_W
=h_W\,\frac{z_\chi^2\,\sin^2\theta \sin ^2\rho 
		 \big( \cos 2 \theta-2\cos ^2\theta \cos 2 \rho -3\big)
		 -4\big(z_\rho\,\sin\theta +z_\theta\,\cos \theta\sin \rho \cos \rho\big)^2}{\mathsf{r}^4 \big(\cos ^2\rho+\sin ^2\theta
		   \sin ^2\rho\big)^3}~.
		   \label{zzT}
\end{align}
Applying the Todorov operator (\ref{todorov}) we can open the indices and easily obtain the explicit
expression of $\big\langle T_{\mu\nu}\big\rangle_W$ in our coordinate system, namely
\begin{equation}
\big\langle T_{\mu\nu}\big\rangle_W=\cD_\mu \cD_\nu\Big(
z^{\lambda} z^{\kappa} \big\langle T_{\lambda\kappa}\big\rangle_W\Big)~.
\end{equation}

For the one-point function of $H_{\mu\nu}$ and $\bar{H}_{\mu\nu}$ we need to rely on the procedure for parity odd quantities given in \cite{Lauria:2018klo}. Adapting it to our case, 
we find two possible structures:
\begin{align}
 z_1^{\mu} z_2^{\nu} \big\langle H_{\mu\nu} +
 \bar{H}_{\mu\nu} \big\rangle_W
 = k_1\, \frac{\e_{IJK}P^I Z_1^J Z_2^K }{(P\circ P)^2}
 +k_2\,\frac{\e_{ABC} P^A Z_1^B Z_2^C}{(P\circ P)^2}~,
 \label{HH0}
\end{align}
where $I,J,K$ run over the orthogonal directions and $A,B,C$ run over the parallel directions.
To determine the constants $k_1$ and $k_2$, we use the supersymmetric Ward identities
that allow us to relate these coefficients to the prefactor $h_W$ appearing in the one-point function
of the stress-energy tensor. This calculation 
is described in Appendix~\ref{app:k1k2h} and the result is
\begin{equation}
k_1=0~,\qquad k_2=\frac{3h_W}{8}~.
\label{k1k2are}
\end{equation}
Inserting this in (\ref{HH0}), we then obtain
\begin{align}
 z_1^{\mu} z_2^{\nu} \big\langle H_{\mu\nu} +\bar{H}_{\mu\nu} \big\rangle_W
 = \frac{3h_W}{8}\,\cos^2 \theta \sin^2 \rho\,\frac{(z_{1 \phi} z_{2 \theta}-z_{2 \phi} z_{1 \theta})\tan \theta+ (z_{1 \rho} z_{2 \phi}-z_{2 \rho} z_{1 \phi})\cot \rho}{\mathsf{r}^3
 \big(\cos ^2\rho+\sin ^2\theta \sin ^2\rho\big)^2}~. 
\end{align}
Opening the indices and projecting onto the self-dual and anti self-dual parts, we find
\begin{equation}
\begin{aligned}
\big\langle H_{\a}{}^{\b}\big\rangle_W \,&\equiv\,  \big\langle H_{\mu\nu}\big\rangle_W (\sigma^{\mu\nu})_{\a}{}^{\b} =\frac{3\ii h_W}{4}\,\frac{\cos\theta \cos\rho \,(\tau^1)_{\a}{}^{\b}
-\sin\theta \,(\tau^2)_{\a}{}^{\b}}{\mathsf{r}^3
 \big(\cos ^2\rho+\sin ^2\theta \sin ^2\rho\big)^2}~,\\[2mm]
\big\langle \bar{H}^{\dot \a}{}_{\dot\b}\big\rangle_W \,&\equiv\,  \big\langle \bar{H}_{\mu\nu}\big\rangle_W (\bar{\sigma}^{\mu\nu})^{\dot \a}{}_{\dot\b}
=-\frac{3\ii h_W}{4}\,\frac{\cos\theta \cos\rho \,(\tau^1)^{\dot\a}{}_{\dot\b}
+\sin\theta \,(\tau^2)^{\dot\a}{}_{\dot\b}}{\mathsf{r}^3
 \big(\cos ^2\rho+\sin ^2\theta \sin ^2\rho\big)^2}~,
\end{aligned}
\label{HHbar}
\end{equation}
where $\tau^i$ are the usual Pauli matrices.

The last one-point function, that of the scalar superprimary operator
$O_2$, is the easiest one. Its functional form can be extracted from \cite{Billo:2016cpy} and, in our
coordinate system, reads
\begin{align}
	\label{vevO2}
		\braket{O_2}_W=\frac{3h_W}{8}\,\frac{1}{P\circ P}=\frac{3h_W}{8}\,
		\frac{1}{\mathsf{r}^2 \big(\cos^2 \rho+ \sin^2 \theta \sin^2 \rho\big)}~.
\end{align}
The coefficient $3h_W /8$ has been fixed from the superconformal Ward identities (see also
\cite{Fiol:2015spa}).

\paragraph{Absence of anomalies:} The functional form of the one-point functions
\eqref{zzT}, \eqref{HHbar} and \eqref{vevO2} on $S^4$ has been obtained from that of the
corresponding one-point functions on $\mathbb{R}^4$ by performing a conformal transformation.
However, this transformation is affected by a Weyl anomaly and thus we have to make sure that
this anomaly will not plague our results. To show this, we can use a simple argument inspired
by \cite{Bianchi:2016xvf}.

Let us recall that the one-point function of the stress-energy tensor on $S^4$ is not vanishing, even in the absence of a defect, and that it contains a contribution proportional to the anomaly coefficient $a$ \cite{Brown:1977sj}\,%
\footnote{For conformally flat manifolds there is no contribution from the B-type anomalies.}. 
For a supersymmetric field theory in the presence of additional background fields, like the $\cN=2$ SYM theory we are considering, the conformal anomaly is constructed out of the full 
Weyl supergravity multiplet and not just out of the background metric \cite{Kuzenko:2013gva,Gomis:2015yaa}. As a consequence, we expect non-vanishing one-point functions for the various components of the stress tensor multiplet. These would all be proportional to the anomaly coefficient $a$. This anomalous contribution is a local feature of the stress tensor multiplet, which is not affected by the presence or absence of a defect. This is very natural since one never expects that bulk CFT data, like the anomaly coefficients, are modified by a defect. 
Therefore, under a Weyl transformation $\widehat{G}_{\mu\nu}\to G_{\mu\nu}=\e^{2\s} \widehat{G}_{\mu\nu}$ of a flat metric $\widehat{G}_{\mu\nu}$, 
the stress tensor one-point function in the presence of a Wilson line $W$ changes as follows
\begin{align}
\frac{\big\langle \widehat{T}_{\mu\nu} W \big\rangle}{\big\langle W \big\rangle}~~\to~~
\frac{\big\langle {T}_{\mu\nu} W \big\rangle}{\big\langle W \big\rangle}
=\rme^{-2\s}\,\frac{\big\langle \widehat{T}_{\mu\nu} W \big\rangle}{\big\langle W \big\rangle}+
\big\langle {T}_{\mu\nu} \big\rangle
\end{align}
where $\widehat{T}_{\mu\nu}$ is the stress tensor in flat space. The
last term in the right hand side is the anomalous contribution, while the
term proportional to $\rme^{-2\s}$ is the result of the conformal transformation
applied to the one-point function in the flat space. In the case where the conformal transformation
maps $\mathbb{R}^4$ to $S^4$, this term is just what we have denoted by 
$\big\langle {T}_{\mu\nu} \big\rangle_W$ in the previous subsection. 
Indeed, from (\ref{XW1}) we have
\begin{align}
\big\langle {T}_{\mu\nu} \big\rangle_W= 
\frac{\big\langle {T}_{\mu\nu} W \big\rangle}{\big\langle W \big\rangle}-
\big\langle {T}_{\mu\nu} \big\rangle=
\rme^{-2\s}\,\frac{\big\langle \widehat{T}_{\mu\nu} W \big\rangle}{\big\langle W \big\rangle}~.
\end{align}

This argument, which applies of course to all other components of the stress tensor multiplet,
shows that the sphere one-point functions that appear in (\ref{dlogW}) are precisely those that are
obtained by performing the conformal transformation on those in flat space, as we have done
to write \eqref{zzT}, \eqref{HHbar} and \eqref{vevO2}. Thus, our result is not affected
by the anomaly. 
Actually, this argument is rather general and holds for an arbitrary line defect in any $\mathcal{N}=2$ SCFT. For the specific case we consider in this paper though, {\it{i.e.}} 
$\mathcal{N}=2$ SYM theory, we know that the anomaly coefficient $a$ does not depend on the coupling and the absence of anomalous contributions can also be ascertained from a simple 
free theory computation.

\subsection{Explicit integration}
We have collected all ingredients that are necessary to perform the integrations in (\ref{dlogW}). 
Let us begin by considering the integral involving the one-point function of the stress-energy tensor.
This has to be regularized by introducing a cutoff $\e$ to keep the integration away from the location of the defect; the result is
\begin{align}
	\label{res1}
		\int \!d^4 \xi \,\sqrt{\det G^0} \,
		\Big[\frac{1}{2}\,\big\langle T_{\mu\nu}\big\rangle_W \, \pa_b G^{\mu\nu}\big|_{b=1}
		\Big]= \Big(\frac{3l^{\prime}-3r^{\prime}
		-3}{\e^3}-\frac{l^{\prime}-r^{\prime}-5}{\e}\Big)2\pi h_W
		+\mathcal{O}(\e)
\end{align}
where
\begin{equation}
l^{\prime}=\partial_b l(b)\big|_{b=1}~,~~~r^{\prime}=\partial_b r(b)\big|_{b=1}
\end{equation}
with $l(b)$ and $r(b)$ being the functions used in (\ref{parametrization}) to parametrize the scales
of the ellipsoid. The expression \eqref{res1} is purely divergent and does not contain any finite contribution. The divergent part is clearly a feature of the regularization procedure since there is no universal logarithmic term. In particular, if we computed the integral \eqref{res1} in dimensional regularization we would simply find zero. For this reason the contribution \eqref{res1} can be discarded.

The other terms in (\ref{dlogW}), instead, yield finite contributions. 
In fact, we find
\begin{subequations}
\begin{align}
\int \!d^4 \xi \,\sqrt{\det G^0} \,
		\Big[16 \,\big\langle H_{\mu\nu}\big\rangle_W
		\,\pa_b \mathsf{T}^{\mu \nu}\big|_{b=1} \Big] &=
		\int \!d^4 x \,\sqrt{\det G} \,
		\Big[\!-2\ii \,\big\langle H_{\a}{}^{\b} \big\rangle_W
		\,\pa_b \mathsf{T}_{\b}{}^{\a}\big|_{b=1}\Big]
		\notag\\
		&=\big(14+4l^{\prime}
		-4r^{\prime}\big)\pi^2 h_W-\frac{3}{2}\pi^4h_W~,\label{resH}\\[2mm]
\int \!d^4 \xi \,\sqrt{\det G^0} \,
		\Big[16 \,\big\langle \bar{H}_{\mu\nu}\big\rangle_W
		\,\pa_b \bar{\mathsf{T}}^{\mu \nu}\big|_{b=1} \Big] &=
		\int \!d^4 \xi \,\sqrt{\det G^0} \,
		\Big[\!-2\ii \,\big\langle \bar{H}^{\dot\a}{}_{\dot\b} \big\rangle_W
		\,\pa_b \bar{\mathsf{T}}^{\dot\b}{}_{\dot\a}\big|_{b=1}\Big]
		\notag\\
		&=\big(14+4l^{\prime}
		-4r^{\prime}\big)\pi^2 h_W-\frac{3}{2}\pi^4h_W~,\label{resHbar}\\[2mm]
\int \!d^4 \xi \,\sqrt{\det G^0} \,
		\Big[\big\langle O_2\big\rangle_W \,\pa_b 
		\widetilde{M}\big|_{b=1}\Big]&=-\big(16+8l^{\prime}
		-8r^{\prime}\big)\pi^2 h_W+3\pi^4h_W~.\label{resO2}
\end{align}
\end{subequations}
It is interesting to observe that, while the individual integrals depend on the constants
$l^{\prime}$ and $r^{\prime}$ that are related to the chosen parametrization of the
ellipsoid scales, remarkably their sum is independent of such a choice. 
Indeed, all terms involving
$l^{\prime}$ and $r^{\prime}$ exactly cancel when we add (\ref{resH}), (\ref{resHbar}) and (\ref{resO2}). Notice that also the terms proportional to $\pi^4$
cancel in the sum. Therefore, discarding the unphysical divergent terms (\ref{res1}) for the aforementioned reasons and 
collecting all the finite contributions, we can rewrite (\ref{dlogW}) as follows
\begin{equation}
\partial_b\ln \big\langle W_b\big\rangle\Big|_{b=1} =12\pi^2 h_W
		+
\frac{\big\langle
\partial_b W_b\big\rangle}{\big\langle W_b\big\rangle}\,\Big|_{b=1}~.
\label{dlogW1}
\end{equation}
The quantity in the left hand side is independent of the parametrization of the ellipsoid, and so also the last term the right hand side must be independent of this parametrization. We can then evaluate it choosing $l(b)=\mathsf{r}/b$, which according to (\ref{parametrization})
implies that $\ell=\mathsf{r}$. In this case the Wilson loop (\ref{WL}) does not explicitly depend on
$b$ and thus $\big\langle \partial_b W_b\big\rangle=0$. 
On the other hand, if we choose a different parametrization for the ellipsoid scales,
we still get this same result. Indeed, as one can see from (\ref{WL}) 
the Wilson loop may explicitly depend on $b$ only through the coefficient $\ell$ in front of the scalar term in the exponent, and the derivative
 $\big\langle \partial_b W_b\big\rangle\big|_{b=1}$ would lead to the integral of a defect one-point function, which clearly vanishes if the defect preserves conformal invariance along its profile. This fact can also be easily checked perturbatively at leading order, as we show in Appendix~\ref{app:check}.

In conclusion the result of our calculation is
\begin{equation}
\partial_b\ln \big\langle W_b\big\rangle\Big|_{b=1} =12\pi^2 h_W~,
\label{dlogW2}
\end{equation}
which proves the conjecture of \cite{Fiol:2015spa}.

\paragraph{Independence on $c_1$, $c_2$ and $c_3$:}
The supergravity background of the ellipsoid given in (\ref{HHsol}) depends on three arbitrary functions $c_1$, $c_2$ and $c_3$ that parametrize the ambiguity in the solution of the Killing spinor equations. These arbitrary functions appear in the $\Delta$-terms given in 
(\ref{DeltaM}) and (\ref{DeltaTs}). However, our result (\ref{dlogW2}) is robust and 
does not depend on these arbitrary functions. Here we would like to explain why this happens.

The $\Delta$-terms in the supergravity background give rise to the following contribution
\begin{equation}
\begin{aligned}
\int \!d^4 \xi \,\sqrt{\det G^0} \,
		\bigg[\!-2\ii \,\big\langle H_{\a}{}^{\b} \big\rangle_W
		\,\pa_b \Delta\mathsf{T}_{\b}{}^{\a}\big|_{b=1}
		-2\ii \,\big\langle \bar{H}^{\dot\a}{}_{\dot\b} \big\rangle_W
		\,\pa_b \Delta\bar{\mathsf{T}}^{\dot\b}{}_{\dot\a}\big|_{b=1}+
		\big\langle O_2\big\rangle_W \,\pa_b 
		\Delta\widetilde{M}\big|_{b=1}
		\bigg]
\end{aligned}
\label{DeltalogW}
\end{equation}
Let us first observe that the terms proportional to $c_i^2$ in $\Delta\widetilde M$ do not contribute
since their $b$-derivative at $b=1$ vanishes because of (\ref{ci}). Similarly, the dependence 
on $c_3$ disappears because in $\pa_b \D  \mathsf{T}_{\a}{}^{\b}$ and 
$\pa_b \D  \bar{\mathsf{T}}^{\dot\a}{}_{\dot\b}$ it multiplies the diagonal matrix $\t^3$, 
while, as one can see from (\ref{HHbar}), the one-point functions 
$\big\langle H_{\a}{}^{\b} \big\rangle_W$ and $\big\langle \bar{H}^{\dot\a}{}_{\dot\b} \big\rangle_W$ are proportional to $\t^1$ and $\t^2$ and hence are anti-diagonal. 

We then remain with the terms proportional to $c_1$ and $c_2$. Evaluating them, we find that
they vanish because they can be recast as total derivatives. Indeed, (\ref{DeltalogW}) becomes
\begin{equation}
\begin{aligned}
3h_W\int \!d^4 \xi \,
		\bigg[\partial_\rho\Big(\frac{\sin\theta \cos\theta\sin^3\rho}{\cos ^2\rho+\sin ^2\theta \sin ^2\rho}\,c_1^\prime\Big)
		+\partial_\theta\Big(\frac{\sin\theta \cos\theta\sin^2\rho}{\cos ^2\rho+\sin ^2\theta \sin ^2\rho}\,c_2^\prime\Big)
		\bigg]=0~.
\end{aligned}
\label{DeltalogW1}
\end{equation}
This proves that the ambiguity in the background solutions does not affect our result (\ref{dlogW2}).

\section{Relating $h_W$ to the emitted energy and the Bremsstrahlung}
\label{Bandcusp}
In Section~\ref{secn:WL} we provided a proof of the relation  (\ref{dlogW2})
between the coefficient $h_W$ of the
stress-energy tensor one-point function and the vacuum expectation value of a half-BPS Wilson loop
on an ellipsoid in the sphere limit.
Here we comment on the connection between $h_W$ and the coefficient appearing in the two-point function of the so-called displacement operator, 
a particular defect excitation related to the breaking of translational invariance
which carries spin one in the space orthogonal to the defect. 

Let us start by considering a conformal Wilson line in four dimensions stretched along one
of the coordinate axes, say for example $x_4$. In this case the displacement operator carries
an index $i=1,2,3$ in the three transverse directions and is denoted by $\mathbb{D}^i$.
Its two-point function is entirely fixed in terms of the defect CFT data and is of the form
\begin{align}
 \big\langle \mathbb{D}^i(x)\, \mathbb{D}^j(0)\big\rangle_W = \frac{C_D \,\d^{ij}}{(x^2)^2}~.
 \label{CD}
\end{align}
The coefficient $C_D$ is a distinctive feature of the CFT and is related to several relevant physical observables. 
For example, $C_D$ determines the small angle limit of the cusp anomalous dimension 
$\G_{\text{cusp}}(\varphi)$, an important quantity which appears as the universal divergent part of a cusped Wilson line expectation value \cite{Correa:2012at}:
\begin{align}
	\label{key}
  		\G_{\text{cusp}}(\varphi)&=-\frac{C_D}{12}\varphi^2 +\mathcal{O}(\varphi^4)
\end{align}
The coefficient of $-\varphi^2$ in this expression is usually called Bremsstrahlung function 
and denoted by $B$; in other words we have
\begin{equation}
C_D=12 B~.
\label{CDB}
\end{equation} 
The same quantity $C_D$ also determines the total energy $\D E_{\text{tot}}$
emitted by an accelerated charged particle \cite{Correa:2012at} under the assumption that 
the initial and final accelerations are equal (and in particular whenever they are equally vanishing, 
{\it{i.e.}} when the particle velocity is asymptotically constant). The formula reads
 \begin{align}
  \D E_{\text{tot}}=\frac{\pi}{6} \,C_D \int \!d\tau \,a^2
  \label{DEis}
 \end{align}
where $a$ is the four-acceleration of the particle and $\tau$ the proper time parametrizing its world-line.

On the other hand, for a particle with four-velocity $u$ and momentum $p$,
one can define another quantity, called the invariant radiation rate 
$\mathcal{R}=u_{\mu} \frac{dp^{\mu}}{d\tau}$. This power rate is not integrated along the world-line, and is manifestly Lorentz invariant. 
Recently in \cite{Fiol:2019woe}, it was found in many different examples of conformal theories 
that $\mathcal{R}$ is always related to $h_W$ in the following simple way
 \begin{align}
  \mathcal{R}=-\frac{16\pi}{3} \,h_W \, a^2~.
  \label{Ris}
 \end{align}
Comparing (\ref{DEis}) and (\ref{Ris}), we can expect a simple relation also
between $C_D$ and $h_W$.
It was already understood in \cite{Lewkowycz:2013laa}, and then formally proven in 
\cite{Bianchi:2018zpb}, that in supersymmetric theories a relation between these two observables 
does indeed exist and can be derived using supersymmetric Ward identities on defect correlation functions. The precise relation is
\begin{align}
 C_D=36 h_W
 \label{CDh}
\end{align}
for any line defect preserving some supersymmetry. 

In general both $C_D$ and $h_W$ are non-trivial functions of the theory parameters (coupling, rank of the gauge group, etc.) and it is remarkable that the relation \eqref{CDh} is theory-independent
and exact. Without supersymmetry, however, there is no universal relation between $C_D$ and $h_W$ \cite{Lewkowycz:2013laa}. To understand why, it is useful to consider the example of free theories. In Table \ref{tab:1} we report the explicit expressions of $C_D$, $h_W$ and their relation for three different simple conformal theories: the free Maxwell theory, the free theory of a conformally coupled scalar and the $\mathcal{N}=2$ U(1) gauge theory. Clearly, the relation between $C_D$ and $h_W$ is not universal and, in general, in the presence of exactly marginal couplings we would expect the proportionality coefficient to depend on these parameters ({\it{i.e.}} that no simple relation exists between the two functions).

\begin{table}[ht]
	\begin{center}
		{\small
		\begin{tabular}{c|c|c|c|c|c}
			\hline
			\hline
			\,CFT \phantom{\bigg|}&  $C_D$ & $h_W$ &$C_D$ vs $h_W$
			&$\mathcal{R}$ & $\mathcal{P}$\\
			\hline
			Maxwell$\phantom{\bigg|}$ & $\frac{e^2}{\pi^2}$ & $\frac{e^2}{32\pi^2}$  & $C_D=32\,h_W$ & $-2\pi\Big(\frac{e^2}{12\pi^2}\Big)\,a^2$
			& 
			$-2\pi\Big(\frac{e^2}{12\pi^2}\Big)\,a^2$\\[7mm]
			Conformal scalar  
			& $\frac{e^2}{2\pi^2}$ & $\frac{e^2}{96\pi^2}$ & $C_D=48\,h_W$&  $-2\pi\Big(\frac{e^2}{36\pi^2}\Big)\,a^2$& 
			$-2\pi\Big(\frac{e^2}{24\pi^2}\Big) \,a^2
-\Big(\frac{e^2}{36\pi}\Big)\,\frac{\dot{a}^0}{\gamma}$ 
			\\[7mm]
			$\mathcal{N}=2$ U(1) & 
			$\frac{3e^2}{2\pi^2}$ & $\frac{e^2}{24\pi^2}$ & $C_D=36\,h_W$&  $-2\pi\Big(\frac{e^2}{9\pi^2}\Big)\,a^2$& 
			$-2\pi\Big(\frac{e^2}{24\pi^2}\Big) \,a^2
-\Big(\frac{e^2}{36\pi}\Big)\,\frac{\dot{a}^0}{\gamma}$ \\[5mm]
			\hline
			\hline
		\end{tabular}
		}
	\end{center}
	\caption{The relevant quantities for three different free conformal theories. The first line refers to the Maxwell theory with Lagrangian $L=\frac{1}{4}F^2$, where $F$ is the electro-magnetic field strength, with a line operator $W=\exp\big(\ii e \int \!dx^\mu A_\mu\big)$. The second line refers to a scalar field $\phi$ with Lagrangian $L=\frac{1}{2}\big(\partial \phi^2+\frac{R}{6}\phi^2\big)$
	where $R$ is the Ricci scalar, with a line operator $W=\exp\big(\ii e \int \!d\tau \,\phi\big)$. The third line refers to $\cN=2$ SYM theory described in Section~\ref{secn:ellipsoid} with gauge group U(1). Notice that the coupling constant $e^2$ used here is related to the Yang-Mills coupling $\gym^2$ used there as $e^2=\gym^2/2$.}
	\label{tab:1}
\end{table}

The other two quantities indicated in Table \ref{tab:1} are the aforementioned invariant radiation rate $\cR$ and the emitted power $\mathcal{P}=\frac{dp^0}{dt}$. 
The relation of the latter with $C_D$ and $h_W$ is subtle since 
$\mathcal{P}$ is not Lorentz invariant and thus is dependent on the observer.
In particular, the expression of $\mathcal{P}$ always contains a Lorentz-invariant 
term proportional to $C_D \,a^2$, 
but it may also contain a boundary term proportional 
to the time-derivative of the time-component of the acceleration $\dot{a}^0$. 
This boundary term may contribute to the integral defining the total emitted energy 
if the initial and final accelerations are not equal \cite{Fiol:2019woe}, thus modifying equation \eqref{DEis}.
This is what happens for example when the acceleration is such that $\dot a^{\mu}=-a^2 u^{\mu}$ with constant $a^2$, which is the configuration considered in \cite{Lewkowycz:2013laa}. This additional term explains the failure in finding a universal relation between the total emitted energy and the stress tensor one-point function, using the argument of \cite{Lewkowycz:2013laa}\footnote{We are grateful to B. Fiol and J. Montoya for a useful discussion on this issue}. 

In presence of supersymmetry, using (\ref{CDh}) and the results 
of Section~\ref{secn:WL}, we can conclude that
for any $\cN=2$ conformal SYM theory the coefficient $C_D$ of the two-point function of the
displacement operator is given by
\begin{equation}
C_D=\frac{3}{\pi^2}\,\partial_b\ln \big\langle W_b\big\rangle\Big|_{b=1}
\label{CDW}
\end{equation}
or, equivalently, that in these theories the Bremsstrahlung function $B$ is
\begin{equation}
B=\frac{1}{4\pi^2}\,\partial_b\ln \big\langle W_b\big\rangle\Big|_{b=1}
\label{BW}
\end{equation}
as conjectured in \cite{Fiol:2015spa}.

\section{Matrix model calculation}
\label{sec::mm}
The relation (\ref{dlogW2}) between the coefficient $h_W$ in the stress tensor one-point function and the $b$-derivative of the ellipsoid Wilson loop, which also implies the relations (\ref{CDW}) and (\ref{BW}) for $C_D$ and $B$, relies on the superconformal symmetry of the gauge theory on the ellipsoid constructed in \cite{Hama:2012bg}. In that same reference, supersymmetric localization was applied to this theory to express its partition function and the expectation value of circular Wilson loops in terms of a matrix model. This makes it possible to explicitly evaluate $h_W$ using matrix model techniques. 

\subsection{$h_W$ in the localization matrix model}
We start by reviewing the $\cN=2$ ellipsoid matrix model obtained in \cite{Hama:2012bg},
which is a generalization of the matrix model for SYM theories on the sphere derived
in \cite{Pestun:2007rz}.
For concreteness, we focus here on the case in which the gauge group is $\mathrm{SU}(N)$, the matter fields transform in a representation $\cR$ such the $\beta$-function vanishes 
and the Wilson loop is in the fundamental representation.
According to the localization principle, the only non-vanishing contributions to the path integrals 
in (\ref{vevO}) and (\ref{Zb}) arise from the following saddle point values of the fields:
\begin{align}
\label{saddle}
	A_\mu=0~, ~~~ \phi=\bar{\phi}=-\frac{\ii}{2}\,a_0~, ~~~\cD_{\cI\cJ}=-\ii \,w_{\cI\cJ}\,a_0~,
\end{align}
where $a_0$ is a $N\times N$ matrix taking values in the $\mathfrak{su}(N)$ Lie Algebra. The explicit expression of $w_{\cI\cJ}$ can be found in \cite{Hama:2012bg}.
The classical action (\ref{Sb}) at this saddle saddle point becomes\,%
\footnote{We denote matrix model quantities by calligraphic letters.} 
\begin{align}
	\label{clsad}
		\cS_b=\frac{8\pi^2}{\gym^2} \ell \tilde\ell \,\tr a_0^2~,
\end{align} 
while the circular BPS Wilson loop (\ref{WL}) becomes
\begin{align}
	\label{wlsp}
		\cW_b = \frac{1}{N}\tr \exp (-2\pi\ell \,a_0)~. 
\end{align}
 
The path integral measure appearing in the partition function and in any other expectation value, reduces to the integration over the matrix $a_0$.
Besides the Gaussian factor arising from $\rme^{-\cS_b}$, the integrand comprises also 
a one-loop determinant, that accounts for the fluctuations around the saddle point, 
and a non-perturbative instanton part. Both of these terms turn out to depend only 
on the ellipsoid scales $\ell$ and $\tilde\ell$ appearing in (\ref{defellipsoid}) and not on $r$. Moreover, the product $\ell\tilde{\ell}$ and the matrix $a_0$ always occur together in the combination $\sqrt{\ell\tilde\ell}\, a_0$. One can thus eliminate entirely the dependence on the product $\ell\tilde\ell$ by changing the integration variable from $a_0$ to the matrix\,%
\footnote{In \cite{Hama:2012bg} the change of variable from $a_0$ to $\hat a_0 = \sqrt{\ell\tilde{\ell}} \,a_0$ is performed. We prefer to rescale $a_0$ also with a factor of $\sqrt{{8\pi^2}/{\gym^2}}$ so that the classical action $\cS_b$
becomes simply $\tr a^2$. This leads to a Gaussian term $\exp(-\tr a^2)$ in the matrix model integrand, while the one-loop determinant and the instanton factor get organized, respectively,
into a perturbative and a non-perturbative expansion in $\gym$. This $\gym$-dependent 
rescaling is the matrix-model equivalent of the rescaling one needs to do on the gauge fields 
to make the coupling constant $\gym$ appear in the covariant derivatives.
The overall minus sign in (\ref{defa}) is irrelevant; we insert it simply because we like to work with a Wilson loop operator in the matrix model with a positive exponent, see (\ref{wlsp}).} 
\begin{align}
	\label{defa}
		a = - \sqrt{\ell\tilde{\ell}} \,\sqrt{\frac{8\pi^2}{\gym^2}}\, a_0~.
\end{align}
The overall constant factors arising from the Jacobian for this change of variable cancel out in all properly normalized expectation values between the integral in the numerator and the partition function in the denominator. 
When written in terms of the matrix $a$, both the one-loop determinant and the instanton terms
only depend on the squashing parameter $b = \sqrt{\ell/\tilde \ell}$, and for $b=1$ they 
reduce to the expressions obtained on the sphere in \cite{Pestun:2007rz}.
Moreover, as shown in \cite{Hama:2012bg}, they are symmetric in 
the exchange $b\leftrightarrow 1/b$. As a consequence of this symmetry, the partition function 
\begin{align}
	\label{partb}
		\cZ_b = \int \! da~\rme^{-\tr a^2} \,\big|\cZ^\text{1-loop}_b\big|^2
		\, \big|\cZ^\text{inst}_b\big|^2
\end{align}
does not depend on $b$ at first order, namely
\begin{align}
	\label{derZb1}
		\partial_b \cZ_b\,\Big|_{b=1} = 0~.
\end{align}

\paragraph{The general expression of $h_W$:}
As we stated above, we want to compute $h_W$ using equation (\ref{conj1}) by evaluating 
the right hand side in the matrix model, namely by 
\begin{align}
	\label{htolog}
	h_W = \frac{1}{12\pi^2}\,\partial_b\log \big\langle \cW_b\big\rangle\Big|_{b=1}~.
\end{align}
In terms of the matrix $a$, the Wilson loop (\ref{wlsp}) reads
\begin{align}
	\label{wlspm}
		\cW_b = \frac{1}{N}\tr \exp \Big(\frac{b\,\gym}{\sqrt{2}}\, a\Big)~,
\end{align}
and its expectation value is
\begin{align}
	\label{derWbmm}
		\big\langle \cW_b\big\rangle = \frac{1}{\cZ_b} \int da~ 
		\cW_b\, \rme^{-\tr a^2} \,\big|\cZ^\text{1-loop}_b\big|^2
		\, \big|\cZ^\text{inst}_b\big|^2~. 
\end{align}
Due to (\ref{derZb1}), in computing $\partial_b \big\langle \cW_b\big\rangle$ at $b=1$ we get a contribution only when the derivative is applied to the operator $\cW_b$ itself. Thus, we obtain
\begin{align}
	\label{derbB}
		\partial_b \ln \big\langle\, \cW_b\,\big\rangle \Big|_{b=1}
		= \frac{\big\langle\partial_b \cW_b\big|_{b=1}\big\rangle\phantom{\Big|}}{\big\langle \cW\big\rangle}\,
		\equiv \,\frac{\big\langle\cW^{\,\prime}\big\rangle}{\big\langle \cW\big\rangle}~.   
\end{align}
Here $\cW$ stands for $\cW_{b=1}$, that is
\begin{equation}
	\label{Wdef}
		\cW = \frac{1}{N}\tr \exp \Big(\frac{\gym}{\sqrt{2}}\, a\Big) = 
		1 + \frac{\gym^2}{4N}  \tr a^2 + O(\gym^3) \ldots~,
\end{equation} 
while 
\begin{align}
	\label{derbW}
		\cW^{\,\prime} =\partial_b \cW_b\big|_{b=1}=  \frac{\gym}{\sqrt{2}} \frac 1N 
		\tr \Big(a\,\exp\Big(\frac{\gym a}{\sqrt{2}}\Big)\Big)
		= \frac{\gym^2}{2N} \tr a^2 + O(\gym^3)~.
\end{align}
Note that we have the identity
\begin{equation}
\cW^{\,\prime} = \gym\,\frac{\partial \cW}{\partial \gym}~.
\label{Wprimeg}
\end{equation}
In (\ref{derbB}), both expectation values in the right hand side
are given by expressions analogous to (\ref{derWbmm})
but at $b=1$, {\it{i.e.}} they are expectation values in the matrix model on the round sphere. 

Inserting (\ref{derbB}) into (\ref{htolog}) expresses $h_W$ in terms of expectation values of operators in the sphere matrix model:
\begin{align}
	\label{fs4mm}
		h_W = \frac{1}{12\pi^2} \frac{\big\langle\cW^{\,\prime}\big\rangle}{\big\langle \cW\big\rangle}~.
\end{align}
Let us observe that in the matrix model it is convenient to choose a strategy, implemented through the rescaling (\ref{defa}), such that the $b$-derivative acts on the operator only. This is the opposite of what happened in the field theory proof of Section \ref{secn:WL}, where the $b$-dependence occurred only through the action.

\paragraph{The $\cN=4$ case:} In the $\cN=4$ SYM theory,  
the matrix model is purely gaussian as both the one-loop determinant and the instanton factor reduce to $1$. Then, after using (\ref{Wprimeg}) in (\ref{fs4mm}), the $\gym$-derivative commutes with the expectation value and thus, as already derived in \cite{Correa:2012nk}, one has
\begin{equation}
	\label{N4B}
		h_W\Big|_{\cN=4}= \frac{1}{12\pi^2}\, \gym \frac{\partial \ln \big\langle\cW\big\rangle }{\partial \gym}~.
\end{equation}
This big simplification no longer occurs in the $\cN=2$ case, due to the non-trivial 1-loop determinant and instanton factors. Nevertheless the quantity in (\ref{derbW}), and then through eq. (\ref{htolog}) the value of $h$ and $B$, can be computed in a standard fashion in the interacting $\cN=2$ matrix model on $S^4$. In particular,  we will employ the techniques of \cite{Billo:2019fbi} to describe its perturbative expansion in $\gym$. 

\subsection{Perturbative expansion}
We now want to explicitly evaluate $h_W$ in a $\cN=2$ superconformal gauge theory using (\ref{fs4mm}).
We consider the perturbative limit in which the coupling $\gym$ is small and the instanton contributions
become trivial, namely we set $\cZ_\tmb{inst}=1$. 
The one-loop determinant can instead be expanded
as follows:
\begin{align}
	\label{interactiveaction}
		|\cZ_{\mathrm{1-loop}}|^2 =~\rme^{-\cS_\text{int}}~,
\end{align}		
where
\begin{align}		
	\label{intact2}
		\cS_{\mathrm{int}}= \sum_{n=2}(-1)^{n}\left(\frac{\gym^2}{8\pi^2}\right)^{n}\,\frac{\zeta(2n-1)}{n}\trp a^{2n}~.
\end{align}
Notice that the absence of the $\gym^2$ term in this expansion is due to the fact that we are considering a conformal theory for which the $\beta$-function vanishes.
In the right hand side of (\ref{intact2}) we used the notation introduced in \cite{Billo:2019fbi}:
\begin{align}
	\label{deftrp}
	\Tr_{\cR}^\prime \bullet \,= \,\Tr_{\cR} \bullet - \Tr_{\mathrm{adj}} \bullet
\end{align}
where $\cR$ is the representation in which the matter hypermultiplets transform.
In the $\cN=4$ SYM theory, where $\cR$ is the adjoint, we easily see that
$\cS_\tmb{int}=0$. 
For $\cN=2$ models, instead, this combination accounts for the matter content 
of the ``difference  theory'' $(\cN=2)-(\cN=4)$, namely the theory in which 
the adjoint hypermultiplets of the $\cN=4$ model are removed and replaced by the matter hypermultiplets in the representation $\cR$ \cite{Andree:2010na}.

The vacuum expectation value of any observable $f$ in the interacting matrix model 
can be expressed in terms of vacuum expectation values computed in the Gaussian matrix model, which we distinguish by a subscript $0$. In particular, we can rewrite (\ref{fs4mm}) as
\begin{align}
	\label{hvev0}
		h_W=\frac{1}{12\pi^2}
	 \frac{\big\langle\cW^{\,\prime}\,\rme^{-\cS_\text{int}}\big\rangle_0}{
	 \big\langle \cW\,\rme^{-\cS_\text{int}}\big\rangle_0}~.
\end{align}
Expanding $\cW$ and $\cW^{\,\prime}$, as well as $\cS_\text{int}$, in series of $\gym$ we obtain the perturbative expansion of $h_W$ in terms of expectation values of multi-traces of powers of the matrix $a$ in the Gaussian model. Such quantities can be easily computed in a recursive way, 
see for instance \cite{Billo:2017glv,Billo:2018oog}, relying on the Wick theorem. If
we write $a= a^c t_c$, where the $\mathfrak{su}(N)$ generators $t_c$ in the fundamental representation are normalized so that $\tr\, t_c t_d = \delta_{cd}/2$, we have\,%
\footnote{We  normalize the flat measure as 
	$da = \prod_{c} \left(da^c/\sqrt{2\pi}\right)$,
	so that  $\int da\, \rme^{-\tr a^2}=1$.
	In this way the contraction (\ref{propabc}) immediately follows.
} 
\begin{align}
	\label{propabc}
		\big\langle a^c a^d\big\rangle_0 = \delta^{cd}~.		
\end{align}
Using such techniques we can compute $h_W$ to any desired perturbative order.

\paragraph{Transcendentality driven expansion:}
It is interesting to organize the computation in terms of the Riemann zeta-values appearing in (\ref{intact2}). Expanding (\ref{hvev0}) in powers of $\gym$, we get an expression of the form
\begin{equation}
\label{hexp}
\begin{aligned}
		h_W & = \gym^2 \,x_1 \big(1 + O(\gym^2)\big) + \gym^6 \,\zeta(3)\, x_3  \big(1 + O(\gym^2)\big)
		+ \gym^8\, \zeta(5)\, x_5 \big(1 + O(\gym^2)\big)\\[1mm]
		& ~~~+ \gym^{10} \Big[\zeta(7)\, x_7 \big(1 + O(\gym^2)\big)+ \zeta(3)^2\, x_{3,3}
		\big(1 + O(\gym^2)\big)\Big] + \ldots
\end{aligned}
\end{equation}
where the coefficients $x_{n_1,n_2,\ldots}$ can be explicitly computed.

Let us then introduce the quantity $\tilde h_W$ obtained by keeping, for each Riemann zeta-value, only the lowest term in $\gym$, namely
\begin{align}
	\label{htexp}
		\tilde h_W & = \gym^2\, x_1  + \gym^6\, \zeta(3)\, x_3 
		+ \gym^8\, \zeta(5)\, x_5
		+ \gym^{10}\, \big[\zeta(7)\, x_7  + \zeta(3)^2\, x_{3,3}\big]
	    + \ldots~.
\end{align}
This quantity is interesting for the comparison with explicit field-theoretic perturbative computations that we will carry out in the next section.

Considering the expression of $h_W$ given in (\ref{hvev0}), we see that it reduces to $\tilde h_W$
if we keep only the lowest term in the perturbative expansions of both $\cW$ and $\cW^{\,\prime}$ given in (\ref{Wdef}) and (\ref{derbW}). Thus we can formally resum (\ref{htexp}) and write  
\begin{align}
	\label{htis}
		\tilde h_W = \frac{1}{12\pi^2}\frac{\gym^2}{2N} 
		\frac{\big\langle\tr a^2\,\rme^{-\cS_\text{int}}\big\rangle_0}{
	 \big\langle \rme^{-\cS_\text{int}}\big\rangle_0}= 
	 \frac{1}{12\pi^2}\frac{\gym^2}{2N} \,\big\langle\tr a^2\big\rangle
\end{align}
to express $\tilde h_W$ in terms of the propagator of the interacting 
matrix model. This latter is given by 
\begin{equation}
	\label{mmprop}
	\big\langle a^c a^d\big\rangle = \delta^{cd}\, \big(1 + \Pi\big)~,
\end{equation}
where $\Pi$ is a $\gym$-dependent constant describing the effect of the perturbative corrections to the propagator. Using this in (\ref{htis}), we find that $\tilde h_W$ is given by
\begin{align}
	\label{htisPi}
		\tilde h_W = \frac{1}{12\pi^2}\,\frac{\gym^2(N^2-1)}{4N} \,\big(1 + \Pi\big)~.
\end{align}
The corrections $\Pi$ were computed in \cite{Billo:2019fbi} with the result\,%
\footnote{In fact, the generic term proportional to a single Riemann zeta value has the expression
\begin{align*}
		(-1)^n\left(\frac{\gym^2}{8\pi^2}\right)^n\zeta(2n-1)\,\,\cC_{2n}^\prime~.	
\end{align*}
 }
\begin{align}
	\label{Pig3}
		\Pi &= \zeta(3) \left(\frac{\gym^2}{8\pi^2}\right)^2 \cC^{\prime}_{4}
		- \zeta(5) \left(\frac{\gym^2}{8\pi^2}\right)^3 \cC^{\prime}_{6} + \cO(g^8)~,
\end{align}
where $\cC_{2n}^\prime$ is the totally symmetric contraction of the tensor
\begin{align}
	\label{defC}
		\cC^\prime_{c_1\ldots c_{2n}} = \trp T_{c_1}\ldots T_{c_{2n}}~. 
\end{align}
In general $\cC_{2n}^\prime$ is a rational function in $N$ (for more details we refer to
Section 3 of \cite{Billo:2019fbi}). For instance, for the conformal SQCD theory (with $N_f=2N$) one finds
\begin{align}
	\label{C4C6qcd}
		\cC_4^\prime=-3(N^2+1)~,~~~
		\cC_6^\prime=-\frac{15(N^2+1)(2N^2-1)}{2 N}~.
\end{align}
Similar expressions can be easily worked out at higher order and for other superconformal theories
with matter fields transforming in different representations.

Exploiting these methods and using the relations (\ref{CDW}) and (\ref{BW}), one can derive the
perturbative expansion of the coefficient $C_D$ in two-point function of the displacement operator and the Bremsstrahlung function $B$, at any desired order. 

\section{Field theory interpretation}
\label{sec:ft}
We now compare the results of the previous sections 
to the computation of the Bremsstrahlung function $B$, of the normalization 
$C_D$ in two-point function of the displacement operator, and of the normalization $h_W$ of the stress-energy one-point function using ordinary perturbative field theory in flat space. 
This comparison is not meant as a check of the relation (\ref{conj1}) of these quantities to the Wilson loop on the ellipsoid, since this is no longer conjectured but proven. Rather, it is meant to illustrate 
how the matrix model results based on this relation suggest how to organize the diagrammatic computations. These suggestions might be useful in the future for studying related quantities and/or different theories. 

We will focus on the lowest order contributions in $\gym$ for each given structure of 
Riemann zeta values. In the matrix model we introduced the notation $\tilde h_W$ for the sum 
of all such contributions to $h_W$ given in (\ref{htexp}) and (\ref{htis}); 
analogously we will use the notations $\tilde B$ and $\tilde C_D$.
As shown in (\ref{htisPi}), in the matrix model $\tilde h_W$ is proportional to the 
propagator. This fact suggests that also on the field-theory side the diagrams contributing to 
$\tilde h_W$, $\tilde B$ and $\tilde C_D$ are given by propagator corrections. 
We will see that for the Bremsstrahlung and for the displacement two-point function this is indeed natural. It is instead much less obvious for the one-point function of operators in the stress-energy multiplet.

\paragraph{Notations and conventions:}
In order to rely on previous literature, we perform a change of conventions with respect to Sections \ref{secn:ellipsoid} and \ref{secn:WL}. We redefine the adjoint scalar fields of the vector multiplet
by
\begin{align}
	\label{redphi}
		\phi \to \frac{\ii \,\gym}{\sqrt{2}} \,\phi~,
		~~~
	\bar\phi \to \frac{\ii\, \gym}{\sqrt{2}} \,\bar\phi~,
\end{align}
while all other components of the gauge multiplet are rescaled by $\gym$, namely 
$A_\mu\to \gym A_\mu$, etc. Having done this, the sum of the YM and matter Lagrangians given in (\ref{LYM}) and (\ref{Lhm}), in flat space and with all supergravity background fields set to zero, reduces to the Lagrangian described -- in $\cN=1$ superfield notation and in the Fermi-Feynman gauge -- in Section 4 of \cite{Billo:2019fbi}. This Lagrangian 
yields canonical (super) propagators.
In particular, at tree level we have
\begin{equation}
\label{propagators}
\begin{aligned}	
		\big\langle A_\mu^c(x)\,A_\nu^d(y)\big\rangle_0 
		&= \delta^{cd}\,\delta_{\mu\nu}\, \Delta(x-y)~,\phantom{\Big|}\\[1mm]
		\big\langle \phi^c(x)\,\bar\phi^{\,d}(y)\big\rangle_0 &
		= \delta^{cd}\,\Delta(x-y)~,
\end{aligned}
\end{equation}
where 
\begin{equation}
	\label{Delta}
		\Delta(x) = \int\!\frac{d^Dk}{(2\pi)^D}\,\frac{\rme^{\ii\,k\cdot x}}{k^2}~.
\end{equation}
with $D=4-2\epsilon$.

\paragraph{Propagator corrections:} In the $\cN=4$ SYM theory, the tree-level propagators (\ref{propagators}) receive no corrections. 
In the $\cN=2$ case, instead, they are corrected in perturbation theory, and take the form
\begin{equation}
\label{propagatorscorr}
\begin{aligned}
		\big\langle A_\mu^c(x)\,A_\nu^d(y)\big\rangle &= (1 + \Pi)\, \delta^{cd}\,\delta_{\mu\nu}\, \Delta(x-y)~,\\[1mm]
		\big\langle \phi^c(x)\,\bar\phi^{\,d}(y)\big\rangle &= (1 + \Pi)\,\delta^{cd}\,\Delta(x-y)~.
\end{aligned}
\end{equation}
In \cite{Billo:2019fbi} it has been argued, and then shown explicitly up to three loops,
that the correction factor $\Pi$ introduced above coincides with the factor $\Pi$ appearing in the  
matrix model given in (\ref{mmprop}). 

\subsection{The Bremsstrahlung function}
\begin{figure}[ht]
	\begin{center}
		\includegraphics[scale=0.7]{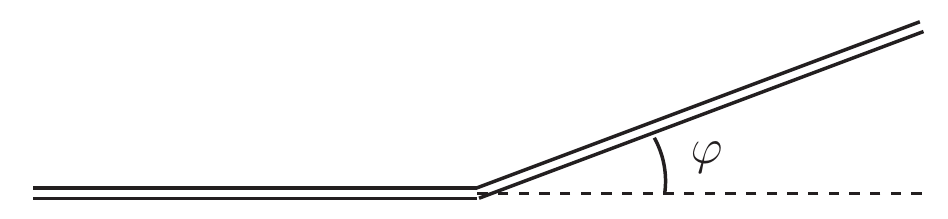}
	\end{center}
	\caption{The contour of a Wilson line with cusp angle $\varphi$.}
	\label{fig:cusp}
\end{figure}

We now compute the leading order coefficient of the small angle expansion of the cusp anomalous dimension (see equation (\ref{key})). This quantity arises from the expectation value of a cusped Wilson line $W_\cusp$ which we take in the fundamental representation of SU($N$). 
Its contour is made of two semi-infinite rays parametrized as follows
\begin{equation}
\begin{aligned}
x^\mu&=v_1^\mu\,\t_1 \quad\mbox{for}\quad-\infty < \t_1 < 0 ~,\\
x^\mu&=v_2^\mu\,\t_2 \quad\mbox{for}\quad 0 < \t_2 < +\infty ~.
\end{aligned}
\label{contour}
\end{equation}
The velocity vectors $v_1^\mu$ and $v_2^\mu$ are such that $v_1\cdot v_1=v_2\cdot v_2=1$. They define the cusp angle\footnote{Note that the angle $\varphi$ of the present section has nothing to do with the ellipsoid coordinate defined in \eqref{coords}} $\varphi$ (see figure~\ref{fig:cusp}) by the relation 
\begin{equation}
 v_1\cdot v_2 = \cos \varphi~.
\label{v1v2a}
\end{equation}

\noindent
The cusped Wilson line is explicitly defined by
\begin{equation}
W_\cusp=\frac{1}{N}\,\tr\mathcal{P} \exp\Big(\gym\!\int_{-\infty}^0\!\!d\t_1\,L_1(\t_1)+
\gym\!\int_0^{+\infty}\!\!d\t_2\,L_2(\t_2)\Big)~,
\label{Wloop}
\end{equation}
where we introduced the generalized connections
\begin{equation}
\label{defL1L2}
\begin{aligned}	
		L_1(\t_1)&= \ii \,v_1\cdot A(v_1\t_1)+\frac{1}{\sqrt{2}}\Big(\rme^{+\ii\,\vartheta/2}\,\phi(v_1\t_1)
		+\rme^{-\ii\,\vartheta/2}\,\bar\phi(v_1\t_1)\Big)~,\\
		L_2(\t_2)&= \ii \,v_2\cdot A(v_2\t_2)+\frac{1}{\sqrt{2}}\Big(\rme^{-\ii\,\vartheta/2}\,\phi(v_2\t_2)
		+\rme^{+\ii\,\vartheta/2}\,\bar\phi(v_2\t_2)\Big)~.
\end{aligned}
\end{equation}
Here $\vartheta$ is an ``internal'' angular parameter that can be defined at the cusp \cite{Drukker:1999zq,Correa:2012nk}; it can be set to zero without any problem.

\begin{figure}[ht]
\vspace{0.3cm}
	\begin{center}
		\includegraphics[scale=0.7]{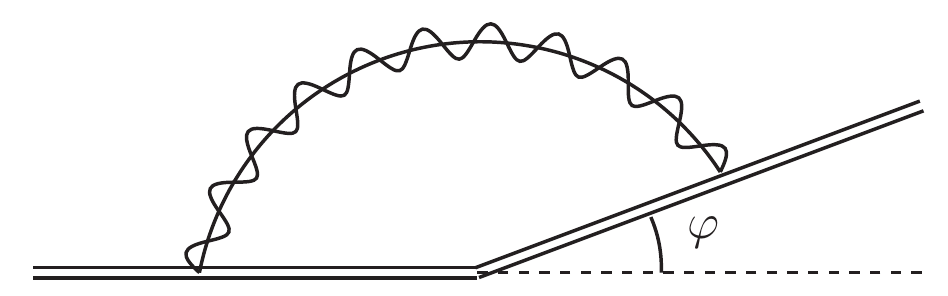}
	\end{center}
	\caption{The $\gym^2$-contribution to the vacuum expectation value of a cusped Wilson line.
		The double straight/wiggled line stands for the sum of the gluon and scalar propagators.}
	\label{fig:cusp1}
\end{figure}

Expanding $W_\cusp$ in $\gym$, we find that its vacuum expectation value at order 
$\gym^2$ is given by the diagram represented in figure~\ref{fig:cusp1}. 

Using the explicit expression of the Wilson line and the propagators \eqref{propagators}, this 
leads to write
\begin{align}
	\label{cWL1}
		\big\langle W_\cusp \big\rangle =
		1 + \,\gym^2\,\frac{N^2-1}{2N}\,\big(\cos\varphi-\cos\vartheta\big)
		\,I(\varphi) + O(\gym^4)~,
\end{align}
where\,%
\footnote{Following \cite{Grozin:2015kna}, we regulate the IR divergence of the $\t_1$ and $\t_2$ 
	integrals by introducing a dumping factor $\rme^{-\ii\delta(\t_1-\t_2)}$ with $\mathrm{Im}\,\delta>0$
	which suppresses the contributions from the large $(-\t_1+\t_2)$ region
	and introduces the dependence on the IR cut-off $\delta$.} 
\begin{equation}
I(\varphi) = \int\!\frac{d^Dk}{(2\pi)^D}\,\frac{1}{k^2\,(k\cdot v_1-\delta)\,(k\cdot v_2-\delta)}~.
\label{Iis}
\end{equation}
This integral is evaluated in Appendix~\ref{app:ft}. Substituting the result (\ref{Int5})
in (\ref{cWL1}), we get
\begin{equation}
\label{cWL2}
\big\langle W_\cusp \big\rangle = 1 - \frac{1}{\varepsilon}\,
\Big(\frac{\gym^2}{8\pi^2}\Big)\,\frac{N^2-1}{2N}\,
\frac{\varphi\,(\cos\varphi-\cos\vartheta)}{\sin\varphi}~.
+O(\gym^4)
\end{equation}
The cusp anomalous dimension $\Gamma_\cusp$ is defined by \cite{Polyakov:1980ca}\,%
\footnote{Often the definition of $\Gamma_\cusp$ is given within a cut-off regularization scheme, in which case $1/(2\epsilon)$ gets replaced by $\log \left(\Lambda_\tmb{UV}/\Lambda_\tmb{IR}\right)$.}
\begin{equation}
	\label{defGcusp}
\big\langle W_\cusp \big\rangle = \exp\Big(\!-\frac{1}{2\epsilon}\,\Gamma_\cusp\Big)~.
\end{equation}
Taking the logarithm of (\ref{cWL2}) and expanding for small angles, we find
\begin{align}
	\label{GtoB}
		\Gamma_\cusp\simeq
		-\big(\varphi^2-\vartheta^2\big)\,B
\end{align}
with
\begin{equation}\label{Bfttree}
B=\Big(\frac{\gym^2}{8\pi^2}\Big)\,\frac{N^2-1}{2N}+O(\gym^4)~.
\end{equation}
This agrees with the lowest order term in the matrix model result (\ref{htisPi}), taking into account that $B= 3h_W$.

\begin{figure}[ht]
	\begin{center}
		\includegraphics[scale=0.83]{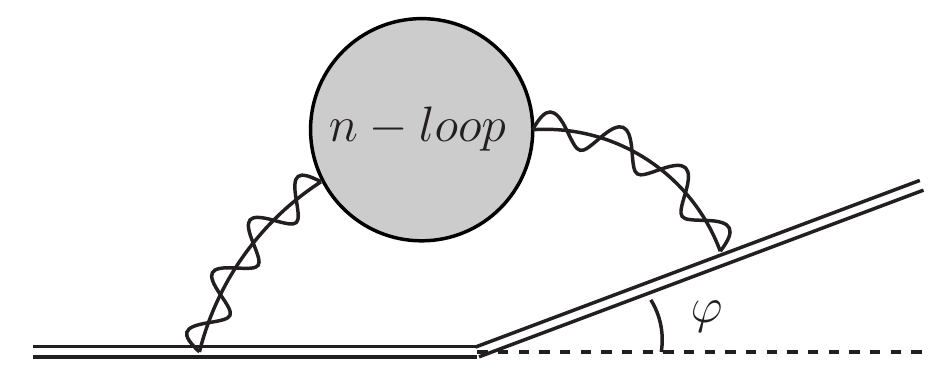}
	\end{center}
	\caption{The contribution to the vacuum expectation value of the cusped Wilson line arising
		from the a single, loop corrected, propagator - of the gluon or of the scalar.}
	\label{fig:cuspn}
\end{figure}

The form of (\ref{htisPi}) indicates that the sum of all perturbative corrections contributing to the lowest order for each transcendentality weight, which we denoted by $\tilde B$, 
can be obtained by replacing in the above derivation the tree level propagators 
(\ref{propagators}) with their loop-corrected counterparts (\ref{propagatorscorr}). In other words, at $n$ loops, we just have to consider the diagram represented in figure~\ref{fig:cuspn}. 

Indeed, it is not difficult to realize that considering diagrams with more propagators attached to the Wilson line increases the order in $\gym$ without giving rise to higher transcendentality. 
The only difference in the explicit expression of the diagrams in figure \ref{fig:cuspn} with respect to the tree-level case of figure \ref{fig:cusp1}, is an overall factor of $(1+\Pi)$. In this way we get
\begin{equation}
	\label{Bftcomplete}
		\tilde B=\Big(\frac{\gym^2}{8\pi^2}\Big)\,\frac{N^2-1}{2N}\,(1+\Pi)~,
\end{equation}
in perfect agreement with (\ref{htisPi}), since $\tilde B = 3 \tilde h_W$.

\subsection{The displacement two-point function}
\label{subsec:disp}
We now consider the field-theory computation of the coefficient $C_D$ of the 
displacement two-point function, introduced in (\ref{CD}).
In \cite{Correa:2012at} this quantity was shown to be related to the Bremsstrahlung function by $C_D=12 B$ in the $\cN=4$ SYM case. This relation holds as well in any $\cN=2$ superconformal theory, and it is understandable at the diagrammatic level in a simple way. 
 
We take a circular Wilson loop\,%
\footnote{We could have chosen as well a straight Wilson line.} in the fundamental representation
given by
\begin{align}
\label{Wline}
W &= \frac{1}{N} \tr \cP \exp \Big(\ii\,\gym \!\int_0^{2\pi}\! \!d\tau L(\tau)
\Big)~,
\end{align}
where
\begin{equation}
L(\tau)=A_\mu \dot{x}^\mu -\ii\, \frac{|\dot{x}|}{\sqrt{2}}(\phi +\bar{\phi})
\end{equation}
with the circular contour being parametrized as $x^\mu(\tau) = (R\cos \tau, R\sin\tau,0,0)$ 
for $\tau\in [0,2\pi]$.
Rather than the displacement operator $\mathbb{D}^i$, in this case 
it is easier to consider its scalar superpartner $\mathbb{O}$. While $\mathbb{D}^i$ arises from the breaking of the conservation of the stress-energy tensor by the Wilson loop defect, the scalar operator $\mathbb{O}$ arises from the breaking of the conservation law for the SO$(1,1)_R$ R-symmetry current. {From} this fact, 
following the prescription in \cite{Bianchi:2018zpb}, one can determine its explicit expression 
finding
\begin{align}
\label{Displscalar}
\mathbb{O}(\tau) = \frac{\ii\, \gym\,R}{\sqrt{2}} \big(\phi(\tau) -\bar{\phi}(\tau)\big)
\end{align}
where $\phi(\tau) \equiv \phi(x(\tau))$ and similarly for $\bar{\phi}$.

The functional form of the defect two-point function of this operator is fixed by the residual conformal symmetry, and its coefficient is related to the one of the displacement two-point function by supersymmetric Ward identities. For the circular Wilson loop we are considering, 
this amounts to 
\begin{align}
	\label{Displ2points}
		\big\langle \mathbb{O}(\tau_1)\,\mathbb{O}(\tau_2)\big\rangle_W 
		= \frac{C_D}{12}\,\frac{1}{(1-\cos\tau_{12})^2}
\end{align}
where $\tau_{12}=\tau_1-\tau_2$.
Using \eqref{Wline} and \eqref{Displscalar}, at the lowest order in $\gym$, we find
\begin{equation}
\label{Displpert}
\begin{aligned}
		\big\langle \mathbb{O}(\tau_1)\,\mathbb{O}(\tau_2)\big\rangle_W &= 
		\frac{1}{N} \tr \cP \,\Big\langle
		\rme^{\ii \,\gym \int_{0}^{\tau_1} \!d\tau L(\tau)}~
		\mathbb{O}(\tau_1) ~\rme^{\ii\, \gym \int_{\tau_1}^{\tau_2} \!d\tau L(\tau)}
		~\mathbb{O}(\tau_2) ~\rme^{\ii\, \gym \int_{\tau_2}^{2\pi}\! d\tau L(\tau)}
		\Big\rangle \\
		&= -\frac{\gym^2 R^2}{4N} \Big\langle
		\big(\phi^c(\tau_1) -\bar{\phi}^c(\tau_1)\big)\,\big(\phi^c(\tau_2) -\bar{\phi}^c(\tau_2)
		\big)	\Big\rangle + O(\gym^4)~.
\end{aligned}
\end{equation}
Using the tree-level scalar propagator (\ref{propagators}) and the explicit parametrization 
$x(\tau)$, we find
\begin{align}
	\label{Displpert2}
		\big\langle \mathbb{O}(\tau_1)\,\mathbb{O}(\tau_2)\big\rangle_W &
		=\frac{\gym^2(N^2-1)}{16\pi^2N} \frac{1}{(1-\cos\tau_{12})^2}+ O(\gym^4)~.
\end{align}
Thus, comparing with (\ref{Displ2points}), we obtain
\begin{align}\label{Cdtree}
C_D=12\, \Big(\frac{\gym^2}{8\pi^2}\Big)\,\frac{N^2-1}{2N}+ O(\gym^4)~,
\end{align}
which agrees with (\ref{Bfttree}) since $C_D = 12 B$. 
This tree-level computation of $C_D$ is based on the insertion of a scalar propagator 
attached to the defect, and is strictly analogous to what we have done in the previous subsection
for the calculation of $B$; the only difference is that in that case both the scalar and the gluon propagator contribute.

\begin{figure}[ht]
	\begin{center}
		\includegraphics[scale=0.83]{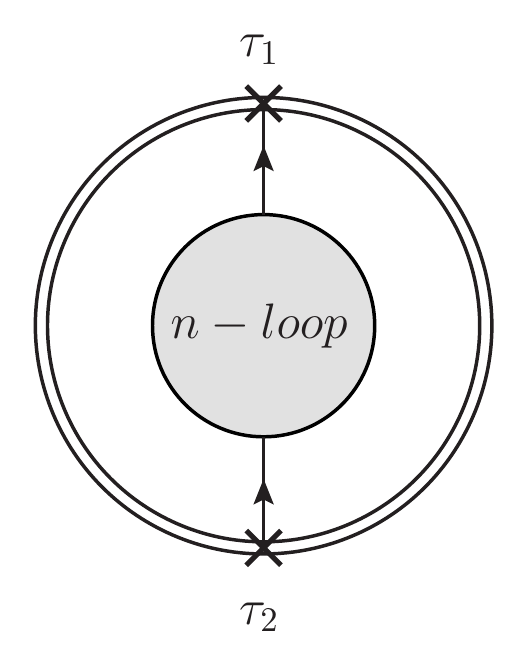}
	\end{center}
	\caption{The contribution to the two-point function of the scalar partner of the displacement operator arising from the $n$-loop correction of the scalar propagators.}
	\label{fig:displacement}
\end{figure}

The matrix model result (\ref{htisPi}) tells us that the contributions at the lowest order for each transcendentality are simply obtained by replacing the tree-level scalar propagator with the full propagator (\ref{propagatorscorr}), as represented in figure~\ref{fig:displacement}.

By summing all these contributions, we produce an extra factor of $(1+\Pi)$ so that
\begin{align}\label{Cdcorr}
\tilde C_D=12\, \Big(\frac{\gym^2}{8\pi^2}\Big)\,\frac{N^2-1}{2N} \,(1+\Pi) = 12 \tilde B~.
\end{align}

\subsection{The stress tensor one-point function}
\label{subsec:sto}
We finally consider the direct diagrammatic computation of the $h_W$ appearing in the defect
one-point functions of the operators of the stress-energy tensor multiplet on the sphere. To do so
we consider the scalar component of this multiplet, namely the operator $O_2$ defined in the
last line of (\ref{varscmultiplet}), which in terms of the rescaled adjoint scalar fields becomes\,%
\footnote{Notice that here we do not include the factor of $\gym$ in the rescaling of $\phi$ and
$\bar{\phi}$, to avoid introducing in the operator an explicit dependence on the coupling constant.}
\begin{equation}\label{O2ft}
O_2 (x)=\tr[\bar \phi \phi] (x)-\frac18 q^{\mathcal{I}} q_{\mathcal{I}}(x)~. 
\end{equation}
As before, we take the defect to be the circular Wilson loop (\ref{Wline}).

The one-point function of $O_2$ in the presence of $W$ is fixed by the conformal symmetry and
depends on the orthogonal scalar product $P\circ P$, as shown in \eqref{vevO2}. 
While in Section~\ref{secn:WL} we used the sphere projection, here we project on $\mathbb{R}^4$. Then, we exploit the residual conformal symmetry to place $O_2$ in the origin, where 
$P\circ P = R^2/4$. In this way we have
\begin{align}
\label{vevO2ft}
\big\langle \,O_2\,\big\rangle_W = \frac{3 h_W}{2 R^2}~.
\end{align}
Using (\ref{Wline}), at the lowest order we find
\begin{align}
\label{O2ft1}
\big\langle \,O_2\,\big\rangle_W&= \frac{\gym^2}{2N} \,\frac{R^2}{2} 
\oint d\tau_1 d\tau_2 \, \Big\langle \, \tr\, [\bar \phi(0) \phi(0)] \,
\tr \!\big[ (\phi +\bar{\phi})(x(\tau_1))\, (\phi +\bar{\phi})(x(\tau_2)) \big]\Big\rangle 
+ O(\gym^4)~.
\end{align}
Inserting the tree-level scalar propagator (\ref{propagators}) and taking into account that 
$x(\tau_i)^2 =R^2$, we get
\begin{align}
\label{O2ft2}
\big\langle \,O_2\,\big\rangle_W&= \frac{\gym^2(N^2-1)}{8N}\frac{1}{4\pi^2 R^2} + O(\gym^4)~, 
\end{align}
from which it follows that 
\begin{equation}
\label{aOtree}
h_W=\frac{1}{3}\, \Big(\frac{\gym^2}{8\pi^2}\Big)\,\frac{N^2-1}{2N}  + O(\gym^4)~, 
\end{equation}
in agreement with the lowest order term in the matrix model result (\ref{htisPi}), and the relations
$C_D=12B=36h_W$.

\begin{figure}[ht]
	\begin{center}
		\includegraphics[scale=0.83]{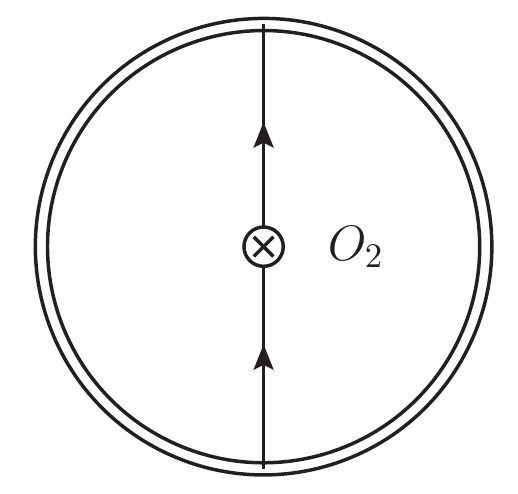}
	\end{center}
	\caption{Tree level contribution to the one-point function of $O_2$}
	\label{fig:O2}
\end{figure}

We note, however, that already at tree level the diagrammatic expansion of this observable 
differs significantly from that of the Bremsstrahlung function $B$ and of the normalization constant $C_D$ in displacement two-point function, because it involves two propagators, and not just one,
as is clear from figure~\ref{fig:O2}. 

Despite this fact, the matrix model result (\ref{htisPi}) for $\tilde h_W$ suggests that the loop diagrams that correct the result at leading order in each transcendentality should organize themselves in terms of loop corrections to a single scalar propagator. This is far from obvious from the point of view of the Feynman diagrams, which are not so easy to compute beyond one loop. 
Indeed, $O_2$ does not belong to the class of chiral operators which enjoy nice cancellation properties due to superconformal symmetry (see for example \cite{Billo:2017glv,Billo:2018oog,Gerchkovitz:2016gxx,Baggio:2014ioa}). In this case, the matrix model could therefore provide non-trivial suggestions on how one should organize the higher loop diagrams 
contributing to the correlators of non-chiral operators. This is an interesting point which is
currently under investigation\,%
\footnote{Work in progress by L. Bianchi, M. Bill\`o, F. Galvagno, P. Gregori and A. Lerda.}.

\vskip 1.5cm
\noindent {\large {\bf Acknowledgments}}
\vskip 0.2cm
We thank Paolo Gregori, Luca Griguolo, Edoardo Lauria, Madalena Lemos, Marco Meineri, Silvia Penati, Igor Pesando and Domenico Seminara for many useful discussions.

\noindent
The work of L.B. is supported by the European Union’s Horizon 2020
research and innovation program under the Marie
Sklodowska-Curie grant agreement No 749909.
The work of M.B.~and A.L. is partially supported by the MIUR PRIN Contract 
2015 MP2CX4 ``Non-perturbative Aspects Of Gauge Theories And Strings''.
The work of A.L. is partially supported also by ``Fondi Ricerca Locale dell'Universit\`a 
del Piemonte Orientale".
\vskip 1cm
\begin{appendix}
\section{Notations and conventions}
\label{app:Notations}
\subsection*{Notations for indices}
\begin{itemize}
\item
$d=4$ vector indices (ellipsoid): $\mu,\nu,\dots = 1,\dots,4$;

\item
$d=4$ vector indices (flat space): $m,n,\dots = 1,\dots,4$;

\item
spatial flat space indices: $i,j = 1,2,3$;

\item
$d=5$ embedding space indices (flat space): $M,N\dots = 1,\dots,5$;

\item
$d=6$ light-cone embedding coordinates: $\mathcal{M},\mathcal{N}= 0,\dots,5$;

\item
$d=6$ ``parallel'' indices: $A,B,C = 0,1,2$, and ``orthogonal'' indices:  $I,J,K = 3,4,5$;

\item
$d=4$ chiral and anti-chiral spinor indices: $\alpha,\beta$ and $\dot{\alpha},\dot{\beta}$;

\item
SU$(2)_R$ symmetry indices: $\mathcal{I},\mathcal{J},\dots = 1,2$;

\item
Sp$(r)$ indices: $\mathcal{A},\mathcal{B}= 1,\dots,2r$;

\item
SU$(N)$ adjoint indices: $c,d,\dots = 1,\dots,N^2-1$.
\end{itemize}

\subsection*{Conventions for traces and spinors}
We denote by $\psi$ a chiral spinor of components $\psi_\alpha$, 
and by $\bar \psi$ an anti-chiral spinor of 
components $\bar{\psi}^{\dot{\alpha}}$.
The spinor indices are raised and lowered with the following rules:
\begin{align}
	\label{raislow}
		\psi^\alpha = \epsilon^{\alpha\beta}\,\psi_\beta~,~~~
		\psi_\alpha = \epsilon_{\alpha\beta}\,\psi^\beta~,~~~
		\bar\psi^{\dot{\alpha}} =\epsilon^{\dot{\alpha}\dot{\beta}}\,\bar\psi_{\dot{\beta}}~,~~~
		\bar\psi_{\dot{\alpha}} =\epsilon_{\dot{\alpha}\dot{\beta}}\,\bar\psi^{\dot{\beta}}~,
\end{align}
where
\begin{equation}
	\label{epsilons}
		\epsilon^{12} = \epsilon^{\dot{1}\dot{2}} =\epsilon_{21} = \epsilon_{\dot{2}\dot{1}} = 1~.
\end{equation}
Contraction rules for undotted/dotted indices are:
\begin{align}
\psi\chi &\equiv \psi^\alpha\,\chi_\alpha ~,~~~~~\bar\psi\bar\chi
 \equiv \bar\psi_{\dot{\alpha}}\,\bar\chi^{\dot{\alpha}}~.
\end{align}
We realize the Euclidean Clifford algebra
\begin{equation}
	\label{cliff4}
		\sigma_m\bar\sigma_n + \sigma_n\bar\sigma_m =
		2\,\delta_{mn}\,\mathbf{1}
\end{equation}
by means of the matrices $(\sigma^m)_{\alpha\dot\beta}$ and
$(\bar\sigma^{m})^{\dot\alpha\beta}$ that can be taken to be
\begin{equation}
	\label{sigmas}
		\sigma^m =
		(-\ii\tau^i,\mathbf{1})~,\qquad
		\bar\sigma^m = (\ii\tau^i,\mathbf{1})~,
\end{equation}
where $\tau^i$ are the ordinary Pauli matrices. They are such that
\begin{equation}
	\label{traspsigma}
		(\bar\sigma^{m})^{\dot\alpha\alpha}=\epsilon^{\alpha\beta}\,\epsilon^{\dot{\alpha}\dot{\beta}}(\sigma^m)_{\beta\dot\beta}~.
\end{equation}
Finally we use:
\begin{equation}
\s_{mn} = \frac{1}{2}(\s_m\bar\s_n-\s_n\bar\s_m)~,~~~\bar\s_{mn} = \frac{1}{2}(\bar\s_m\s_n-\bar\s_n\s_m)~,
\end{equation}
where $\s_{mn}$ is anti self-dual, while $\bar\s_{mn}$ is self-dual.

Our conventions for traces over the group generators are as follows. In any representation $\cR$ we
take
\begin{equation}
\tr_{\cR} T^c T^d = i_\cR\,\delta^{cd}
\end{equation}
where $i_\cR$ is the index of $\cR$. In particular, for the fundamental representation of SU($N$), 
we have
\begin{equation}
\tr T^c T^d = \frac{1}{2}\,\delta^{cd}~.
\end{equation}

\section{SUSY transformations}
\label{app:SUSYtransf}
Let us start by listing the on-shell SUSY transformations of the fields in the vector multiplet.
We follow \cite{Hama:2012bg}, but consider the SUSY parameters $\xi$ as Grassmann odd.
\begin{equation}
\label{susygauge}
\begin{aligned}	
		\d A_{\mu}&=\ii\xi^{\mathcal{I}} \s_{\mu} \bar \l_{\mathcal{I}}-\ii\bar \xi^{\mathcal{I}} \bar \s_{\mu} \l_{\mathcal{I}}~,\\
		\d \phi &= - \ii \xi^{\mathcal{I}} \l_{\mathcal{I}}~,\\
		\d \bar \phi &= + \ii \bar \xi^{\mathcal{I}} \bar \l_{\mathcal{I}}~,\\
		\d \l_{\mathcal{I}}&=\frac12 \s^{\mu\nu}\xi_{\mathcal{I}} (F_{\mu\nu}+8\bar \phi T_{\mu\nu}) + 2\s^{\mu} \bar \xi_{\mathcal{I}} D_{\mu} \phi+\s^{\mu} D_{\mu} \bar \xi_{\mathcal{I}} \phi+ 2\ii \xi_{\mathcal{I}} [\phi,\bar \phi]~,\\
		\d \bar \l_{\mathcal{I}}&=\frac12 \bar \s^{\mu\nu}\bar \xi_{\mathcal{I}} (F_{\mu\nu}+8 \phi \bar T_{\mu\nu}) + 2\bar \s^{\mu} \bar \xi_{\mathcal{I}} D_{\mu} \bar \phi+\bar \s^{\mu} D_{\mu}  \xi_{\mathcal{I}} \bar \phi- 2\ii \bar \xi_{\mathcal{I}} [\phi,\bar \phi]~.
\end{aligned}
\end{equation}
This algebra closes on the following field equations
\begin{align}
	\label{eomgauge}
		\bar \s^{\mu}D_{\mu} \l_{\mathcal{I}}=2\ii[\phi,\bar \l_{\mathcal{I}}]~,~~~  
		\s^{\mu}D_{\mu} \bar \l_{\mathcal{I}}=2\ii[\bar\phi, \l_{\mathcal{I}}]~.
\end{align}
For the hypermultiplet the on-shell SUSY transformations are 
\begin{align}
	\label{susyhyper}
		\d q_{\mathcal{I}}&=-\ii\xi_{\mathcal{I}} \psi + \ii \bar \xi_{\mathcal{I}} \bar \psi~,\notag\\
		\d \psi &=2 \s^{\mu} \bar \xi_{\mathcal{I}} D_{\mu} q^{\mathcal{I}}+ \s^{\mu} D_{\mu} \bar \xi_{\mathcal{I}} q^{\mathcal{I}}-4\ii \xi_{\mathcal{I}} \bar \phi q^{\mathcal{I}}~,\notag\\
		\d \bar \psi &=2 \bar \s^{\mu} \xi_{\mathcal{I}} D_{\mu} q^{\mathcal{I}}+ \bar \s^{\mu} D_{\mu}  \xi_{\mathcal{I}} q^{\mathcal{I}}-4\ii \bar\xi_{\mathcal{I}}  \phi q^{\mathcal{I}}~.
\end{align}

Now we consider the stress tensor multiplet. In flat space, the on-shell SUSY transformations are
\begingroup
\allowdisplaybreaks
\begin{align}
	\label{susystm}
		\d O_2&=\ii\bar \chi_{\dot \a \mathcal{I}} \bar \xi^{\dot \a \mathcal{I}} +\ii \xi_{\mathcal{I}}^{\a} \chi_{\a}^{\mathcal{I}}~,\notag\\
		\d\chi_{\a}^{\mathcal{I}}&=H_{\a}{}^{\b} \xi^{\mathcal{I}}_\b+\frac12 j_{\a \dot \a} \bar \xi^{\dot \a \mathcal{I}}+\frac12 t_{\a \dot \a \mathcal{J}}{}^{\mathcal{I}} \bar \xi^{\dot \a \mathcal{J}}+\pa_{\a \dot \a} O_2 \bar \xi^{\dot \a \mathcal{I}}~,\notag\\
		\d\bar \chi_{\dot \a \mathcal{I}}&=-\bar H^{\dot \b}{}_{\dot \a} \bar \xi_{\dot \b \mathcal{I}}+\frac12 j_{\a \dot \a} \xi_{\mathcal{I}}^{\a}+\frac12 t_{\a \dot \a \mathcal{I}}{}^{\mathcal{J}}  \xi_\mathcal{J}^{\a}-\pa_{\a \dot \a} O_2 \xi_{\mathcal{I}}^{\a}~,\notag\\
		\d H_{\a}{}^{\b}&=\frac{\ii}{2} J_{\a \dot \a}{}^{\b}{}_{\mathcal{I}} \bar \xi^{\dot \a \mathcal{I}}+\frac{2 \ii}{3} \big(\pa_{\a \dot \a} \chi^\b_{\mathcal{I}} +\pa^{\b}{}_{\dot \a} \chi_{\a \mathcal{I}}\big)\bar \xi^{\dot \a \mathcal{I}}~,\notag\\
		\d \bar H^{\dot \b}{}_{\dot \a}&=-\frac{\ii}{2} \bar J_{\a \dot \a}{}^{\dot \b \mathcal{I}} \xi^{\a}_{\mathcal{I}}-\frac{2 \ii}{3} \big(\pa_{\a \dot \a} \bar \chi^{\dot \b \mathcal{I}} +\pa_{\a}{}^{\dot \b} \bar \chi_{\dot \a}^{\mathcal{I}}\big) \,\xi_{\mathcal{I}}^{\a}~,\notag\\
		\d j_{\a \dot \a}&=- \frac{\ii}{2} J_{\a \dot \a \b}{}^{\mathcal{I}} \xi_{\mathcal{I}}^{\b} -\frac{\ii}{2} \bar J_{\a \dot \a \dot \b \mathcal{I}} \bar \xi^{\dot \b \mathcal{I}}+\frac{4\ii}{3} \xi_{\mathcal{I}}^\b \big(2 \pa_{\b \dot \a} \chi_{\a}^{\mathcal{I}} -\pa_{\a \dot \a} \chi_{\b}^{\mathcal{I}}\big)+\frac{4\ii}{3} \bar\xi^{\dot \b \mathcal{I}} \big(2 \pa_{\a \dot \b} \bar\chi_{\dot \a \mathcal{I}} -\pa_{\a \dot \a} \bar \chi_{\dot \b \mathcal{I}}\big)~,\notag\\
		\d t_{\a \dot \a \mathcal{I}}{}^{\mathcal{J}}&=\ii J_{\a \dot \a \b}{}^{\mathcal{J}} \xi_{\mathcal{I}}^\b+\ii \bar J_{\a \dot \a \b \mathcal{I}} \bar \xi^{\dot \b \mathcal{J}}+\frac{4\ii}{3} \xi_{\mathcal{I}}^\b \big(2 \pa_{\b \dot \a} \chi_{\a}^{\mathcal{J}} -\pa_{\a \dot \a} \chi_{\b}^{\mathcal{J}}\big)+\frac{4\ii}{3} \bar\xi^{\dot \b \mathcal{J}} \big(2 \pa_{\a \dot \b} \bar\chi_{\dot \a \mathcal{I}} -\pa_{\a \dot \a} \bar \chi_{\dot \b \mathcal{I}}\big)\notag\\
		-\frac12&\d^{\mathcal{J}}_{\mathcal{I}} \Big[\ii J_{\a \dot \a \b}{}^{\mathcal{K}} \xi_{\mathcal{K}}^\b+\ii \bar J_{\a \dot \a \b {\mathcal{K}}} \bar \xi^{\dot \b {\mathcal{K}}}+\frac{4\ii}{3} \xi_{\mathcal{K}}^\b \big(2 \pa_{\b \dot \a} \chi_{\a}^{\mathcal{K}} -\pa_{\a \dot \a} \chi_{\b}^{\mathcal{K}}\big)+\frac{4\ii}{3} \bar\xi^{\dot \b {\mathcal{K}}} \big(2 \pa_{\a \dot \b} \bar\chi_{\dot \a {\mathcal{K}}} -\pa_{\a \dot \a} \bar \chi_{\dot \b {\mathcal{K}}}\big)\Big]~,\notag\\
		\d J_{\a \dot \a \b}{}^{\mathcal{I}}&=2 T_{\a \dot \a \b \dot \b} \bar \xi^{\dot \b \mathcal{I}}  +\frac23 \big(\pa_{\a \dot \a} H_{\b}{}^{\g}+\pa_{\b \dot \a} H_{\a}{}^{\g}\big)\xi^{\mathcal{I}}_\g-2 \pa_{\g \dot \a} H_{\b}{}^{\g} \xi_{\a}^{\mathcal{I}}-2 \pa_{\g \dot \a} H_{\a}{}^{\g} \xi_{\b}^{\mathcal{I}}\notag\\
		&-\bar \xi^{\dot \b \mathcal{I}} \Big(\frac23 \pa_{\a \dot \a} j_{\beta \dot \beta}- \frac13 \pa_{\b \dot \a} j_{\a \dot \b}- \pa_{\a \dot \b} j_{\b \dot \a}\Big) +2\bar \xi^{\dot \b \mathcal{J}} \Big(\frac23 \pa_{\a \dot \a} t_{\beta \dot \beta \mathcal{J}}{}^{\mathcal{I}}- \frac13 \pa_{\b \dot \a} t_{\a \dot \b \mathcal{J}}{}^{\mathcal{I}}- \pa_{\a \dot \b} t_{\b \dot \a \mathcal{J}}{}^{\mathcal{I}}\Big)~,\notag\\
		\d \bar J_{\a \dot \a \dot \b \mathcal{I}}&=-2 T_{\a \dot \a \b \dot \b} \xi_{\mathcal{I}}^{ \b}  -\frac23 \big(\pa_{\a \dot \a}\bar  H^{\dot \g}{}_{\dot \b}+\pa_{\a \dot \b} \bar H^{\dot \g}{}_{\dot\a}\big)\bar \xi_{\dot \g \mathcal{I}}+2 \pa_{\a \dot \g} \bar H_{\dot \b}{}^{\dot \g} \bar \xi_{\dot \a \mathcal{I}}+2 \pa_{\a \dot \g} \bar H_{\dot \a}{}^{\dot \g} \bar \xi_{\dot \b \mathcal{I}}\notag\\
		&- \xi_{\mathcal{I}}^{\b} \Big(\frac23 \pa_{\a \dot \a} j_{\beta \dot \beta}- \frac13 \pa_{\a \dot \b} j_{\b \dot \a}- \pa_{\b \dot \a} j_{\a \dot \b}\Big) +2 \xi_\mathcal{J}^{\b} \Big(\frac23 \pa_{\a \dot \a} t_{\beta \dot \beta \mathcal{I}}{}^{\mathcal{J}}- \frac13 \pa_{\a \dot \b} t_{\b \dot \a \mathcal{I}}{}^{\mathcal{J}}- \pa_{\b \dot \a} t_{\a \dot \b \mathcal{I}}{}^{\mathcal{J}}\Big)~,\notag\\
		\d T_{\a \dot \a \b \dot \b}&=\frac{\ii}4 \xi_{\mathcal{I}}^\g \big(2\pa_{\g \dot \a} J_{\b \dot \b \a}{}^{\mathcal{I}}-\pa_{\a \dot \a} J_{\b \dot \b \g}{}^{\mathcal{I}}\big)-\frac{\ii}4 \bar \xi^{\dot \g \mathcal{I}} \big(2\pa_{\a \dot \g} \bar J_{\b \dot \b \dot \a \mathcal{I}}-\pa_{\a \dot \a} \bar J_{\b \dot \b \dot \g \mathcal{I}}\big) + \big( \{\a,\dot \a\} \leftrightarrow \{\b,\dot \b\} \big)~.
\end{align}
\endgroup
These transformations obey the commutation relations
\begin{align}
	\label{susyalg}
		\Big[\big[\d_{\xi_1}, \d_{\xi_2}\big] , \bullet\Big]
		=-2\ii (\xi_{1 \mathcal{K}}^\a \bar \xi_2^{\dot \a \mathcal{K}}-\xi_{2 \mathcal{K}}^\a \bar \xi_1^{\dot \a \mathcal{K}}) \pa_{\a \dot \a} \bullet~.
\end{align}
It is possible to verify that the normalization factors of the operators listed in (\ref{varscmultiplet})
are consistent with these SUSY transformations.

\section{One-point function of $H_{\a}{}^{\b}$ and $\bar H^{\dot \a}{}_{\dot \b}$
from Ward identities}
\label{app:k1k2h}

Here we show how to fix the coefficients $k_1$ and $k_2$ appearing in \eqref{HH0}, in terms of $h_W$. We do this by using superconformal Ward identities as in 
\cite{Fiol:2015spa,Bianchi:2018zpb}.

Since the relation between $k_1$, $k_2$ and $h_W$ does not depend on the Poincar\'e section, we choose the simplest set-up. Namely we choose flat-space and a straight Wilson line along one
of the coordinate axes, say $x^4$. Then, the projection on the flat-space Poincar\'e section
is defined by the embedding vectors
\begin{align}\
	\label{flatproj}
		P^{\cM}=\Big(\frac{1+ x^2}{2},x^{m}, \frac{1- x^2}{2}\Big)~,~~~
		Z_k^{\cM}=\big(z_k\cdot x,x^{m}, -z_k\cdot x\big)~.
\end{align}
In particular, the components orthogonal to the line defect are $P^i=x^i$ with $i=1,2,3$. For this
projection, the one-point function of the scalar operator $O_2$, given in \eqref{vevO2},
takes the form:
\begin{equation}\label{vevO2flat}
\vev{O_2}_W= \frac{3 h_W}{8} \frac{1}{P\circ P} = \frac{3 h_W}{8} \frac{1}{x_i x^i}~.
\end{equation}
Applying the (flat-space) SUSY transformations given in Appendix~\ref{app:SUSYtransf}, one finds
that the one-point functions of $H_{\a}{}^{\b}$ and $\bar H^{\dot \a}{}_{\dot \b}$ take the
form
\begin{align}
	\label{Hflatsusy}
	\big\langle H_{\a}{}^{\b}\big\rangle_W=-\frac{3\ii h_W}{4 \big(x_i x^i)^2}
	 \, \big(x_i \t^i\big)_\a{}^\b~,~~~
		\big\langle \bar H^{\dot \a}{}_{\dot \b}\big\rangle_W=\frac{3\ii h_W}{4 \big(x_i x^i)^2}
		\, \big(x_i \t^i\big)^{\dot\a}{}_{\dot \b}~.
\end{align}
On the other hand, taking the general form \eqref{HH0}, using the flat-space Poincar\'e section
(\ref{flatproj}, and extracting the components $H_{\a}{}^{\b}$ and $\bar H^{\dot \a}{}_{\dot \b}$,
 we get
\begin{align}
\label{Hflat}
\big\langle H_{\a}{}^{\b}\big\rangle_W=-\frac{2\ii (k_2-k_1) }{\big(x_i x^i)^2}
	 \, \big(x_i \t^i\big)_\a{}^\b~,~~~
		\big\langle \bar H^{\dot \a}{}_{\dot \b}\big\rangle_W=\frac{2\ii (k_2+k_1)}{4 \big(x_i x^i)^2}
		\, \big(x_i \t^i\big)^{\dot\a}{}_{\dot \b}~.
\end{align}
Comparing these expressions with (\ref{Hflatsusy}), we obtain
\begin{equation}
k_1=0~,\qquad k_2=\frac{3h_W}{8}
\label{k1k2are1}
\end{equation}
which is the condition reported in \eqref{k1k2are}.

\section{Tree level computation of $\big\langle\partial_b W_b\big\rangle\big|_{b=1}$}
\label{app:check}
In this appendix we check that $\big\langle\partial_b W_b\big\rangle\big|_{b=1}=0$ at 
leading order in $\gym$.
Starting from the Wilson loop expression \eqref{WL} and considering the parametrization \eqref{parametrization}, after the rescaling (\ref{redphi}) we have
\begin{equation}
\label{dbWL}
\big\langle\partial_b W_b\big\rangle\big|_{b=1} = \frac{\ell'(1)}{d_\mathcal{R}}\,\Tr_{\mathcal{R}}\mathcal{P}\,\Big\langle 
\frac{r\,\gym}{\sqrt{2}}\!\int_\mathcal{C}	\!d\varphi_1\,(\phi+\bar\phi)~
\exp\bigg[\int_\mathcal{C}	\!d\varphi_2\,\Big(\ii\,
A_\varphi+\frac{r\,\gym}{\sqrt{2}}(\phi+\bar\phi)\Big)\bigg]
\Big\rangle~.
\end{equation}
The vacuum expectation value in the right hand side of \eqref{dbWL} is taken on the sphere, where the Wilson loop is placed on the equator. The tree-level term comes from expanding the exponential
at linear order and then from using the tree level propagator of the scalar fields. From Section 5 of \cite{Billo:2019job}, we read that the scalar propagator on the sphere in $D=4-2\epsilon$ dimensions is
\begin{equation}
\big\langle \phi^c(x_1)\,\bar\phi^d(x_2)\big\rangle=\Delta_S(x_{12}) \,\delta^{cd}
\end{equation}
where
\begin{align}
\Delta_S(x_{12})&=\frac{\Gamma(1-\epsilon)}{4\pi(\pi x_{12}^2)^{1-\epsilon}}~.
\end{align}
Since a generic point on the equator is parametrized as $x(\varphi)= r (\cos \varphi, \sin\varphi,0,0)$, one can see that the tree-level term in \eqref{dbWL} is proportional to the following
integral
\begin{equation}
\int_0^{2\pi} \!d\varphi \,\frac{r^2 \Gamma(1-\epsilon)}{4\pi\big(2\pi r^2 (1-\cos \varphi)^2\big)^{1-\epsilon}} = -\frac{2^{2\epsilon-3}\pi^{\epsilon-\frac{1}{2}}r^{2\epsilon}\sec(\pi\epsilon)\Gamma(1-\epsilon)}{\Gamma(\frac{3}{2}-\epsilon)\Gamma(\epsilon)}=O(\epsilon)~.
\end{equation}
This shows that when $\epsilon\to 0$ the tree-level term of $\big\langle\partial_b W_b\big\rangle\big|_{b=1}=0$ vanishes for any parametrization of the ellipsoid scales, 
in agreement with the general remarks outlined in Section~\ref{secn:WL}.

\section{Useful formul\ae\, for the field theory computations}
\label{app:ft}
In the following we will make use of the following integrals:
\begin{itemize}
\item Feynman parametrizations:
\begin{subequations}
\begin{align}
\frac{1}{A^\alpha\,B^\beta}&=\frac{\Gamma(\alpha+\beta)}{\Gamma(\alpha)\,\Gamma(\beta)}\int_0^1 \!dx\,\frac{x^{\alpha-1}(1-x)^{\beta-1}}{\big(x A+(1-x)B\big)^{\alpha+\beta}}
\label{Fx}\\[2mm]
\frac{1}{A^\alpha\,B^\beta}&=
\frac{\Gamma(\alpha+\beta)}{\Gamma(\alpha)\,\Gamma(\beta)}
\int_0^\infty \!dy \,\frac{y^{\beta-1}}{\big(A+y B\big)^{\alpha+\beta}}
\label{Fy}
\end{align}
\end{subequations}
\item The one-loop momentum integral (with Euclidean signature):
\begin{equation}
\int\!\frac{d^Dq}{(2\pi)^D}\,\frac{1}{\big(q^2+M^2\big)^n}=
\frac{\Gamma\big(n-\frac{D}{2}\big)}
{(4\pi)^\frac{D}{2}\,\Gamma(n)}\,\big(M^2\big)^{\frac{D}{2}-n}
\label{intq}
\end{equation}
\item The integral:
\begin{equation}
\int_0^\infty\!dy\,y^\alpha(A y+ B)^\beta =
\frac{\Gamma(-\alpha-\beta-1)\Gamma(\alpha+1)}{\Gamma(-\beta)}\, \frac{B^{\alpha+\beta+1}}{A^{\alpha+1}}~.
\label{Inty}
\end{equation}
\end{itemize}
With these ingredients, we can now perform the calculation of the following integral
\begin{equation}
I(\varphi) = \int\!\frac{d^Dk}{(2\pi)^D}\,\frac{1}{k^2\,(k\cdot v_1-\delta)\,(k\cdot v_2-\delta)}
\label{Iis1}
\end{equation}
where $D=4-2\varepsilon$, and $v_1$ and $v_2$ are two 4-vectors such that
\begin{equation}
v_1\cdot v_1=v_2\cdot v_2=1\quad\mbox{and}\quad v_1\cdot v_2 = \cos \varphi~.
\label{v1v2}
\end{equation}
We follow essentially the procedure outlined in \cite{Grozin:1992yq} (correcting a few typos).

We first use the Feynman parametrization (\ref{Fx})
to combine the two factors that are linear in $k$, obtaining
\begin{align}
I(\varphi)
=\int_0^1\!dx\int\!\frac{d^Dk}{(2\pi)^D}\,\frac{1}{k^2\,\big[\big(x v_1+(1-x) v_2\big)\cdot k-\delta\big]^2}~.
\end{align}
Then, we use the alternative Feynman parametrization (\ref{Fy}) and get
\begin{align}
I(\varphi)&=\int_0^1\!dx\int_0^\infty\!dy
\int\!\frac{d^Dk}{(2\pi)^D}\,\frac{2y}{\big[k^2+ y\big(x v_1+(1-x) v_2\big)\cdot k-y\delta\big]^3}
\label{I1}
\end{align}
Evaluating the integral over $k$, we obtain
\begin{align}
I(\varphi)&=2\int_0^1\!dx\int_0^\infty\!dy\,y
\int\!\frac{d^Dq}{(2\pi)^D}\,\frac{1}{(q^2+M^2)^3}
\label{I2}
\end{align}
with
\begin{align}
M^2
=-y\Big[\frac{y}{4}\big(x^2+(1-x)^2+2x(1-x)\cos\varphi\big)+\delta\Big]~.
\end{align}
Now we can use (\ref{intq}) and get
\begin{align}
I(\varphi)
=-(-1)^{-\varepsilon}\,\frac{\Gamma(1+\varepsilon)}{(4\pi)^{2-2\varepsilon}}
\int_0^1\!dx\int_0^\infty\!dy\,y^{-\varepsilon}\Big[\frac{y}{4}\big(x^2+(1-x)^2+2x(1-x)\cos\varphi\big)+\delta\Big]^{-1-\varepsilon}~.
\end{align}
The integral over $y$ can be computed using (\ref{Inty}), and the result is
\begin{align}
I(\varphi)
=-(-1)^{-\varepsilon}\,\frac{\Gamma(2\varepsilon)\,\Gamma(1-\varepsilon)\,\delta^{-2\varepsilon}}{(2\pi)^{2-2\varepsilon}}
\int_0^1\!dx\,
\frac{1}{\big(x^2+(1-x)^2+2x(1-x)\cos\varphi\big)
^{1-\varepsilon}}~.
\label{Int3}
\end{align}
{From} this expression we explicitly see the UV divergence signaled by the pole for 
$\varepsilon\to 0$.
Since we are ultimately interested in the coefficient of this divergence, we have
\begin{align}
I(\varphi)&=\frac{1}{\varepsilon}\left[-\frac{1}{8\pi^2}
\int_0^1\!dx\,
\frac{1}{\big(x^2+(1-x)^2+2x(1-x)\cos\varphi\big)}\right]+ O(\varepsilon^0)~.
\label{Int4}
\end{align}
The integral over $x$ can be evaluated by setting
\begin{equation}
x=\frac{1}{2}\Big(1+\cot\frac{\varphi}{2}\,z\Big)~.
\end{equation}
In this way we find
\begin{equation}
I(\varphi)=\frac{1}{\varepsilon}\left(-\frac{1}{8\pi^2}\,\frac{\varphi}{\sin\varphi}\right)+ O(\varepsilon^0)~.
\label{Int5}
\end{equation}

\end{appendix}

\providecommand{\href}[2]{#2}\begingroup\raggedright\endgroup


\end{document}